\documentclass[twocolumn, floatfix, resetfootnote]{aastex7}
\usepackage{tikz,amsmath}
\usepackage{color,soul}
\usepackage[english]{babel}
\usepackage{blindtext}
\usepackage{longtable}
\usepackage{multirow}
\usepackage{comment}
\usepackage{float}
\usepackage{hyperref}
\usepackage{physics}
\usepackage{textcomp, gensymb}
\usepackage{subcaption} 
\usepackage{tabularx}

\newcommand{\lgl}{$\log L_{\rm [OII]}$}

\newcommand{\lglg}{$\log L_{\rm [OII]} - \log M_*$}
\renewcommand{\arcsec}{$^{\prime\prime}$}

\begin{document}

\title{\bf DESI EMISSION-LINE GALAXIES: CLUSTERING DEPENDENCE ON STELLAR MASS AND [OII] LUMINOSITY}


\correspondingauthor{Tyler Hagen, Kyle Dawson, Zheng Zheng}

\author[0009-0007-2936-1124,gname=Tyler,sname=Hagen]{T.~Hagen}
\affiliation{Department of Physics and Astronomy, The University of Utah, 115 South 1400 East, Salt Lake City, UT 84112, USA}
\email{tyler.hagen@utah.edu}

\author[0000-0002-0553-3805]{K.~S.~Dawson}
\affiliation{Department of Physics and Astronomy, The University of Utah, 115 South 1400 East, Salt Lake City, UT 84112, USA}
\email{kdawson@astro.utah.edu}

\author[0000-0003-1887-6732]{Z.~Zheng}
\affiliation{Department of Physics and Astronomy, The University of Utah, 115 South 1400 East, Salt Lake City, UT 84112, USA}
\email{zhengzheng@astro.utah.edu}

\author{J.~Aguilar}
\affiliation{Lawrence Berkeley National Laboratory, 1 Cyclotron Road, Berkeley, CA 94720, USA}
\email{jaguilar@lbl.gov}

\author[0000-0001-6098-7247]{S.~Ahlen}
\affiliation{Department of Physics, Boston University, 590 Commonwealth Avenue, Boston, MA 02215 USA}
\email{ahlen@bu.edu}

\author[0000-0001-5537-4710]{S.~BenZvi}
\affiliation{Department of Physics \& Astronomy, University of Rochester, 206 Bausch and Lomb Hall, P.O. Box 270171, Rochester, NY 14627-0171, USA}
\email{sbenzvi@ur.rochester.edu}

\author[0000-0001-9712-0006]{D.~Bianchi}
\affiliation{Dipartimento di Fisica ``Aldo Pontremoli'', Universit\`a degli Studi di Milano, Via Celoria 16, I-20133 Milano, Italy}
\affiliation{INAF-Osservatorio Astronomico di Brera, Via Brera 28, 20122 Milano, Italy}
\email{davide.bianchi1@unimi.it}

\author{D.~Brooks}
\affiliation{Department of Physics \& Astronomy, University College London, Gower Street, London, WC1E 6BT, UK}
\email{david.brooks@ucl.ac.uk}

\author[0000-0001-7316-4573]{F.~J.~Castander}
\affiliation{Institut d'Estudis Espacials de Catalunya (IEEC), c/ Esteve Terradas 1, Edifici RDIT, Campus PMT-UPC, 08860 Castelldefels, Spain}
\affiliation{Institute of Space Sciences, ICE-CSIC, Campus UAB, Carrer de Can Magrans s/n, 08913 Bellaterra, Barcelona, Spain}
\email{fjc@ice.csic.es}

\author{T.~Claybaugh}
\affiliation{Lawrence Berkeley National Laboratory, 1 Cyclotron Road, Berkeley, CA 94720, USA}
\email{tmclaybaugh@lbl.gov}

\author[0000-0002-2169-0595]{A.~Cuceu}
\affiliation{Lawrence Berkeley National Laboratory, 1 Cyclotron Road, Berkeley, CA 94720, USA}
\email{acuceu@lbl.gov}

\author[0000-0002-1769-1640]{A.~de la Macorra}
\affiliation{Instituto de F\'{\i}sica, Universidad Nacional Aut\'{o}noma de M\'{e}xico,  Circuito de la Investigaci\'{o}n Cient\'{\i}fica, Ciudad Universitaria, Cd. de M\'{e}xico  C.~P.~04510,  M\'{e}xico}
\email{macorra@fisica.unam.mx}

\author{P.~Doel}
\affiliation{Department of Physics \& Astronomy, University College London, Gower Street, London, WC1E 6BT, UK}
\email{apd@star.ucl.ac.uk}

\author[0000-0003-4992-7854]{S.~Ferraro}
\affiliation{Lawrence Berkeley National Laboratory, 1 Cyclotron Road, Berkeley, CA 94720, USA}
\affiliation{University of California, Berkeley, 110 Sproul Hall \#5800 Berkeley, CA 94720, USA}
\email{sferraro@lbl.gov}

\author[0000-0002-3033-7312]{A.~Font-Ribera}
\affiliation{Institut de F\'{i}sica d’Altes Energies (IFAE), The Barcelona Institute of Science and Technology, Edifici Cn, Campus UAB, 08193, Bellaterra (Barcelona), Spain}
\email{afont@ifae.es}

\author[0000-0002-2890-3725]{J.~E.~Forero-Romero}
\affiliation{Departamento de F\'isica, Universidad de los Andes, Cra. 1 No. 18A-10, Edificio Ip, CP 111711, Bogot\'a, Colombia}
\affiliation{Observatorio Astron\'omico, Universidad de los Andes, Cra. 1 No. 18A-10, Edificio H, CP 111711 Bogot\'a, Colombia}
\email{je.forero@uniandes.edu.co}

\author{E.~Gaztañaga}
\affiliation{Institut d'Estudis Espacials de Catalunya (IEEC), c/ Esteve Terradas 1, Edifici RDIT, Campus PMT-UPC, 08860 Castelldefels, Spain}
\affiliation{Institute of Cosmology and Gravitation, University of Portsmouth, Dennis Sciama Building, Portsmouth, PO1 3FX, UK}
\affiliation{Institute of Space Sciences, ICE-CSIC, Campus UAB, Carrer de Can Magrans s/n, 08913 Bellaterra, Barcelona, Spain}
\email{gazta@ice.cat}

\author[0000-0003-3142-233X]{S.~Gontcho A Gontcho}
\affiliation{Lawrence Berkeley National Laboratory, 1 Cyclotron Road, Berkeley, CA 94720, USA}
\email{satyagontcho@lbl.gov}

\author[0000-0001-9938-2755]{V.~Gonzalez-Perez}
\affiliation{Centro de Investigaci\'{o}n Avanzada en F\'{\i}sica Fundamental (CIAFF), Facultad de Ciencias, Universidad Aut\'{o}noma de Madrid, ES-28049 Madrid, Spain}
\affiliation{Instituto de F\'{\i}sica Te\'{o}rica (IFT) UAM/CSIC, Universidad Aut\'{o}noma de Madrid, Cantoblanco, E-28049, Madrid, Spain}
\email{violeta.gonzalez@uam.es}

\author{G.~Gutierrez}
\affiliation{Fermi National Accelerator Laboratory, PO Box 500, Batavia, IL 60510, USA}
\email{gaston@fnal.gov}

\author[0000-0003-1197-0902]{C.~Hahn}
\affiliation{Steward Observatory, University of Arizona, 933 N. Cherry Avenue, Tucson, AZ 85721, USA}
\affiliation{Steward Observatory, University of Arizona, 933 N. Cherry Avenue, Tucson, AZ 85721, USA}
\email{chhahn@arizona.edu}

\author[0000-0002-6550-2023]{K.~Honscheid}
\affiliation{Center for Cosmology and AstroParticle Physics, The Ohio State University, 191 West Woodruff Avenue, Columbus, OH 43210, USA}
\affiliation{Department of Physics, The Ohio State University, 191 West Woodruff Avenue, Columbus, OH 43210, USA}
\affiliation{The Ohio State University, Columbus, 43210 OH, USA}
\email{kh@physics.osu.edu}

\author[0000-0002-6024-466X]{M.~Ishak}
\affiliation{Department of Physics, The University of Texas at Dallas, 800 W. Campbell Rd., Richardson, TX 75080, USA}
\email{mishak@utdallas.edu}

\author[0000-0002-0000-2394]{S.~Juneau}
\affiliation{NSF NOIRLab, 950 N. Cherry Ave., Tucson, AZ 85719, USA}
\email{stephanie.juneau@noirlab.edu}

\author{R.~Kehoe}
\affiliation{Department of Physics, Southern Methodist University, 3215 Daniel Avenue, Dallas, TX 75275, USA}
\email{kehoe@physics.smu.edu}

\author[0000-0003-3510-7134]{T.~Kisner}
\affiliation{Lawrence Berkeley National Laboratory, 1 Cyclotron Road, Berkeley, CA 94720, USA}
\email{tskisner@lbl.gov}

\author[0000-0001-6356-7424]{A.~Kremin}
\affiliation{Lawrence Berkeley National Laboratory, 1 Cyclotron Road, Berkeley, CA 94720, USA}
\email{akremin@lbl.gov}

\author[0000-0002-6731-9329]{C.~Lamman}
\affiliation{Center for Astrophysics $|$ Harvard \& Smithsonian, 60 Garden Street, Cambridge, MA 02138, USA}
\email{clamman@g.harvard.edu}

\author[0000-0003-1838-8528]{M.~Landriau}
\affiliation{Lawrence Berkeley National Laboratory, 1 Cyclotron Road, Berkeley, CA 94720, USA}
\email{mlandriau@lbl.gov}

\author[0000-0001-7178-8868]{L.~Le~Guillou}
\affiliation{Sorbonne Universit\'{e}, CNRS/IN2P3, Laboratoire de Physique Nucl\'{e}aire et de Hautes Energies (LPNHE), FR-75005 Paris, France}
\email{llg@lpnhe.in2p3.fr}

\author[0000-0002-3677-3617]{A.~Leauthaud}
\affiliation{Department of Astronomy and Astrophysics, UCO/Lick Observatory, University of California, 1156 High Street, Santa Cruz, CA 95064, USA}
\affiliation{Department of Astronomy and Astrophysics, University of California, Santa Cruz, 1156 High Street, Santa Cruz, CA 95065, USA}
\email{alexie@ucsc.edu}

\author[0000-0003-1887-1018]{M.~E.~Levi}
\affiliation{Lawrence Berkeley National Laboratory, 1 Cyclotron Road, Berkeley, CA 94720, USA}
\email{melevi@lbl.gov}

\author[0000-0003-4962-8934]{M.~Manera}
\affiliation{Departament de F\'{i}sica, Serra H\'{u}nter, Universitat Aut\`{o}noma de Barcelona, 08193 Bellaterra (Barcelona), Spain}
\affiliation{Institut de F\'{i}sica d’Altes Energies (IFAE), The Barcelona Institute of Science and Technology, Edifici Cn, Campus UAB, 08193, Bellaterra (Barcelona), Spain}
\email{mmanera@ifae.es}

\author[0000-0002-1125-7384]{A.~Meisner}
\affiliation{NSF NOIRLab, 950 N. Cherry Ave., Tucson, AZ 85719, USA}
\email{aaron.meisner@noirlab.edu}

\author{R.~Miquel}
\affiliation{Instituci\'{o} Catalana de Recerca i Estudis Avan\c{c}ats, Passeig de Llu\'{\i}s Companys, 23, 08010 Barcelona, Spain}
\affiliation{Institut de F\'{i}sica d’Altes Energies (IFAE), The Barcelona Institute of Science and Technology, Edifici Cn, Campus UAB, 08193, Bellaterra (Barcelona), Spain}
\email{rmiquel@ifae.es}

\author[0000-0002-2733-4559]{J.~Moustakas}
\affiliation{Department of Physics and Astronomy, Siena College, 515 Loudon Road, Loudonville, NY 12211, USA}
\email{jmoustakas@siena.edu}

\author[0000-0001-9070-3102]{S.~Nadathur}
\affiliation{Institute of Cosmology and Gravitation, University of Portsmouth, Dennis Sciama Building, Portsmouth, PO1 3FX, UK}
\email{seshadri.nadathur@port.ac.uk}

\author[0000-0003-3188-784X]{N.~Palanque-Delabrouille}
\affiliation{IRFU, CEA, Universit\'{e} Paris-Saclay, F-91191 Gif-sur-Yvette, France}
\affiliation{Lawrence Berkeley National Laboratory, 1 Cyclotron Road, Berkeley, CA 94720, USA}
\email{npalanque-delabrouille@lbl.gov}

\author[0000-0001-7145-8674]{F.~Prada}
\affiliation{Instituto de Astrof\'{i}sica de Andaluc\'{i}a (CSIC), Glorieta de la Astronom\'{i}a, s/n, E-18008 Granada, Spain}
\email{fprada@iaa.es}

\author[0000-0001-6979-0125]{I.~P\'erez-R\`afols}
\affiliation{Departament de F\'isica, EEBE, Universitat Polit\`ecnica de Catalunya, c/Eduard Maristany 10, 08930 Barcelona, Spain}
\email{ignasi.perez.rafols@upc.edu}

\author[0000-0002-7522-9083]{A.~J.~Ross}
\affiliation{Center for Cosmology and AstroParticle Physics, The Ohio State University, 191 West Woodruff Avenue, Columbus, OH 43210, USA}
\affiliation{Department of Astronomy, The Ohio State University, 4055 McPherson Laboratory, 140 W 18th Avenue, Columbus, OH 43210, USA}
\affiliation{The Ohio State University, Columbus, 43210 OH, USA}
\email{ross.1333@osu.edu}

\author{G.~Rossi}
\affiliation{Department of Physics and Astronomy, Sejong University, 209 Neungdong-ro, Gwangjin-gu, Seoul 05006, Republic of Korea}
\email{graziano@sejong.ac.kr}

\author[0000-0002-6186-5476]{S.~Saito}
\affiliation{Institute for Multi-messenger Astrophysics and Cosmology, Department of Physics, Missouri University of Science and Technology, 1315 N Pine St, Rolla, MO 65409 U.S.A.}
\email{saitos@mst.edu}

\author[0000-0002-9646-8198]{E.~Sanchez}
\affiliation{CIEMAT, Avenida Complutense 40, E-28040 Madrid, Spain}
\email{eusebio.sanchez@ciemat.es}

\author{D.~Schlegel}
\affiliation{Lawrence Berkeley National Laboratory, 1 Cyclotron Road, Berkeley, CA 94720, USA}
\email{djschlegel@lbl.gov}

\author{M.~Schubnell}
\affiliation{Department of Physics, University of Michigan, 450 Church Street, Ann Arbor, MI 48109, USA}
\affiliation{University of Michigan, 500 S. State Street, Ann Arbor, MI 48109, USA}
\email{schubnel@umich.edu}

\author[0000-0002-3461-0320]{J.~Silber}
\affiliation{Lawrence Berkeley National Laboratory, 1 Cyclotron Road, Berkeley, CA 94720, USA}
\email{jhsilber@lbl.gov}

\author{D.~Sprayberry}
\affiliation{NSF NOIRLab, 950 N. Cherry Ave., Tucson, AZ 85719, USA}
\email{david.sprayberry@noirlab.edu}

\author[0000-0003-1704-0781]{G.~Tarl\'{e}}
\affiliation{University of Michigan, 500 S. State Street, Ann Arbor, MI 48109, USA}
\email{gtarle@umich.edu}

\author{B.~A.~Weaver}
\affiliation{NSF NOIRLab, 950 N. Cherry Ave., Tucson, AZ 85719, USA}
\email{benjamin.weaver@noirlab.edu}

\author[0000-0001-5381-4372]{R.~Zhou}
\affiliation{Lawrence Berkeley National Laboratory, 1 Cyclotron Road, Berkeley, CA 94720, USA}
\email{rongpuzhou@lbl.gov}

\author[0000-0002-6684-3997]{H.~Zou}
\affiliation{National Astronomical Observatories, Chinese Academy of Sciences, A20 Datun Road, Chaoyang District, Beijing, 100101, P.~R.~China}
\email{zouhu@nao.cas.cn}

\date{\today}

\begin{abstract}
We measure the projected two-point correlation functions of emission-line galaxies (ELGs) from the Dark Energy Spectroscopic Instrument One-Percent Survey and model their dependence on stellar mass and [OII] luminosity.
We select $\sim$180,000 ELGs with redshifts of $0.8 < z < 1.6$ and define 27 samples according to cuts in redshift and both galaxy properties.
Following a framework that describes the conditional [OII] luminosity-stellar mass distribution as a function of halo mass, we simultaneously model the clustering measurements of all samples at fixed redshift.
Based on the modeling result, most ELGs in our samples are classified as central galaxies, residing in halos of a narrow mass range with a typical median of $\sim$10$^{12.2-12.4}$ $h^{-1} M_\odot$.
We observe a weak dependence of clustering amplitude on stellar mass, which is reflected in the model constraints and is likely a consequence of the 0.5 dex measurement uncertainty in the stellar mass estimates.
The model shows a trend between galaxy bias and [OII] luminosity at high redshift ($1.2 < z < 1.6$) that is otherwise absent at lower redshifts.
\end{abstract}


\section{Introduction}\label{sec:intro}
In our understanding of cosmology, dark matter dominates the total matter distribution.
Primordial fluctuations in the early Universe result in aggregation of dark matter that collapses into dense regions known as halos.
Through the cooling and condensation of gas, galaxies form and evolve within these dark matter halos over cosmic time \citep{white78}.
While the large-scale structure (LSS) of dark matter halos has historically been understood analytically (e.g., \citealt{press74, mo96, zentner07}) and through $N$-body simulations (e.g., \citealt{lacey94, jenkins01}), baryonic physics complicates the role of galaxies in this paradigm and makes galaxies a biased tracer of the underlying matter distribution.
An informative description of this galaxy bias is the relationship between galaxies, their host halos, and the physical properties of both.
This relationship remains unclear and an important topic known as the \emph{galaxy-halo connection}.
A deeper understanding of the galaxy-halo connection can provide insights into galaxy formation, evolution, and the nature of dark matter itself (see, e.g., the review by \citealt{wechsler18}).

One popular route for studying the galaxy-halo connection is through the measurement and modeling of galaxy clustering.
Namely, the two-point correlation function (2PCF) quantifies the excess probability (with respect to a random distribution) of observing two galaxies at a given separation \citep{limber53, limber54, rubin54}.
The clustering of galaxies has been measured extensively across spectroscopic surveys and has revealed dependencies on both directly observable and inferred properties, including color, luminosity, stellar mass, and star formation rate (e.g., \citealt{zehavi11, mostek13, xu18, clontz22}).
Further, many studies have shown that the clustering of halos themselves is also property dependent.
The mass of the halo is most often considered, but secondary properties such as age, concentration, and formation time are also relevant (e.g., \citealt{mo96, gao05, wechsler06, xu18_assembly}).

Many empirical models have been developed to place galaxies in halos according to these properties and describe the galaxy-halo connection.
One such method is known as (subhalo) abundance matching (SHAM), in which galaxies are ordered according to some property--typically stellar mass--and placed in halos ranked according to halo mass or a similar proxy \citep{conroy06}.
Other methods, such as the halo occupation distribution (HOD) formalism, describe the probability distribution of a galaxy sample within halos as a function of halo mass \citep{jing98, seljak00, berlind02}.
The halo occupation statistics can then be studied across samples of varying galaxy properties.
Conceptually similar, the conditional luminosity function (CLF) framework describes the luminosity distribution of galaxies within halos \citep{yang03}.
Each of these occupation models, in conjunction with halo clustering statistics, can be used to fit property-dependent galaxy clustering and place constraints on the galaxy-halo connection \citep{wechsler18}.

In this work, we adopt the conditional color-magnitude distribution (CCMD) framework introduced by \cite{xu18} and adapted by \cite{clontz22} to study the galaxy-halo connection of a sample of star-forming galaxies.
The CCMD model describes the distribution of both galaxy color and luminosity as a function of halo mass and can be considered an extension of the HOD/CLF framework.
With this global parameterization, the occupation statistics and 2PCF of any sample in the color-luminosity plane can be computed.
\cite{clontz22} extend this formalism to describe, at fixed halo mass, the stellar mass-H$\alpha$ luminosity distribution.
Here, we use the same formalism, but for stellar mass and [OII] luminosity (a limited proxy for star formation rate; see \citealt{moustakas06, favole20, favole24}), to study the clustering of emission-line galaxies (ELGs) in the Dark Energy Spectroscopic Instrument (DESI; \citealt{desi16i, desi16ii}) survey.

From a galaxy evolution perspective, ELGs are particularly interesting to study.
They are actively undergoing star formation, providing a direct probe into this important stage of stellar mass growth.
Many models of galaxy formation and evolution have been developed to examine ELGs and their cosmological context (e.g., \citealt{gonzalez-perez20}).
For star-forming galaxies like ELGs, stellar mass and star formation rate are major observable properties that describe their evolutionary status.
Inference of the connection between these properties and the host dark matter halos from galaxy clustering data informs the role of halo environment in shaping such properties and provides an empirical test for galaxy formation models.

Newly formed stars within ELGs can ionize the surrounding gas, producing nebular emission lines that enable an accurate determination of their spectroscopic redshifts.
For this reason, and due to their high number density across redshifts of interest, ELGs are being leveraged in modern spectroscopic surveys to study the matter-dominated epoch of the Universe at $z \sim 1$ \citep{comparat13}.
Specifically, the data from DESI's survey validation program provide more than 300,000 ELGs for our analysis \citep{raichoor23}.

The clustering and galaxy-halo connection of ELGs from DESI's survey validation have been studied individually \citep{rocher23, ortega-martinez24} and in conjunction with other tracers \citep{gao23, gao24, prada25, yu24, yuan25}.
Beyond clustering, the intrinsic properties of this sample have been compared to reference star-forming galaxies \citep{lan24}.
ELG clustering dependence on stellar mass or [OII] luminosity has been studied previously in other surveys \citep{okumura21, gao22}; however, no study in DESI has jointly probed the dependence on both properties as presented here.
Doing so will provide a more detailed and complete description of the galaxy-halo connection, enabling a deeper understanding of galaxy formation.
It will also inform the creation of realistic mock catalogs that can be used to calibrate analysis pipelines and test cosmological models.

This paper is organized as follows.
In Section \ref{sec:data}, we describe the DESI data and our ELG samples.
In Section \ref{sec:clustering}, we present the details of our clustering measurements.
The model is described in Section \ref{sec:model}.
Results are presented and discussed in Section \ref{sec:results}.
We conclude and outline future efforts in Section \ref{sec:conclusions}.

Throughout this work, we adopt a flat Lambda cold dark matter ($\Lambda$CDM) cosmology consistent with \cite{planck18}, using $\Omega_{m,0}=0.310$, $\Omega_{b,0}=0.049$, $H_0 = 100h$ km Mpc$^{-1}$ s$^{-1}$, $h=0.677$, $\sigma_8 = 0.810$, and $n_s = 0.967$.

\section{DESI Data}\label{sec:data}
In this section, we present an overview of the DESI survey validation program, from which we construct ELG samples according to stellar mass and [OII] luminosity.
These galaxy properties are estimated from the DESI spectra and broadband photometry using \textsc{FastSpecFit}\footnote{\url{https://github.com/desihub/fastspecfit}} and taken from the survey validation value-added catalogs\footnote{\url{https://data.desi.lbl.gov/public/edr/vac/edr/fastspecfit/fuji/v3.2/}} \citep{moustakas23}.
Additionally, we characterize the stellar mass measurement uncertainties and provide a custom aperture correction to the [OII] luminosity.

\subsection{One-Percent Survey} \label{subsec:1percent}
Upon completion of its 5 yr operation, the DESI survey will have covered $\sim$14,000 deg$^2$ and measured accurate redshifts for nearly 40 million galaxies and quasars \citep{desi16i,desi16ii}.
To improve over its predecessors, DESI utilizes 5000 robotic fiber positioners distributed over 10 spectrographs to simultaneously collect 5000 spectra \citep{miller24, poppett24}.
The instrument is sensitive from 360 to 980 nm and is attached to the 4 m Mayall telescope at Kitt Peak National Observatory \citep{desi22}.

The parent sample from which DESI targets are selected is provided by the DESI Legacy Imaging Surveys \citep{dey19}, which combine $g$, $r$, and $z$ optical imaging from the Dark Energy Camera \citep{flaugher15} Legacy Survey (or DECaLS), the Beijing-Arizona Sky Survey (or BASS; \citealt{zou17}), and the Mayall $z$-band Legacy Survey (or MzLS; \citealt{dey16}) with mid-infrared $W1$, $W2$, $W3$, and $W4$ imaging from the Wide-field Infrared Survey Explorer (or WISE; \citealt{wright10}).
With this photometry, DESI defines four main extragalactic target classes: the bright galaxy sample \citep{hahn23}, luminous red galaxies \citep{zhou23}, ELGs \citep{raichoor23}, and quasars \citep{chaussidon23}.
Stars in the Milky Way are also observed \citep{cooper23}.

\begin{figure*}
\begin{center}
    \includegraphics[width=0.9\textwidth]{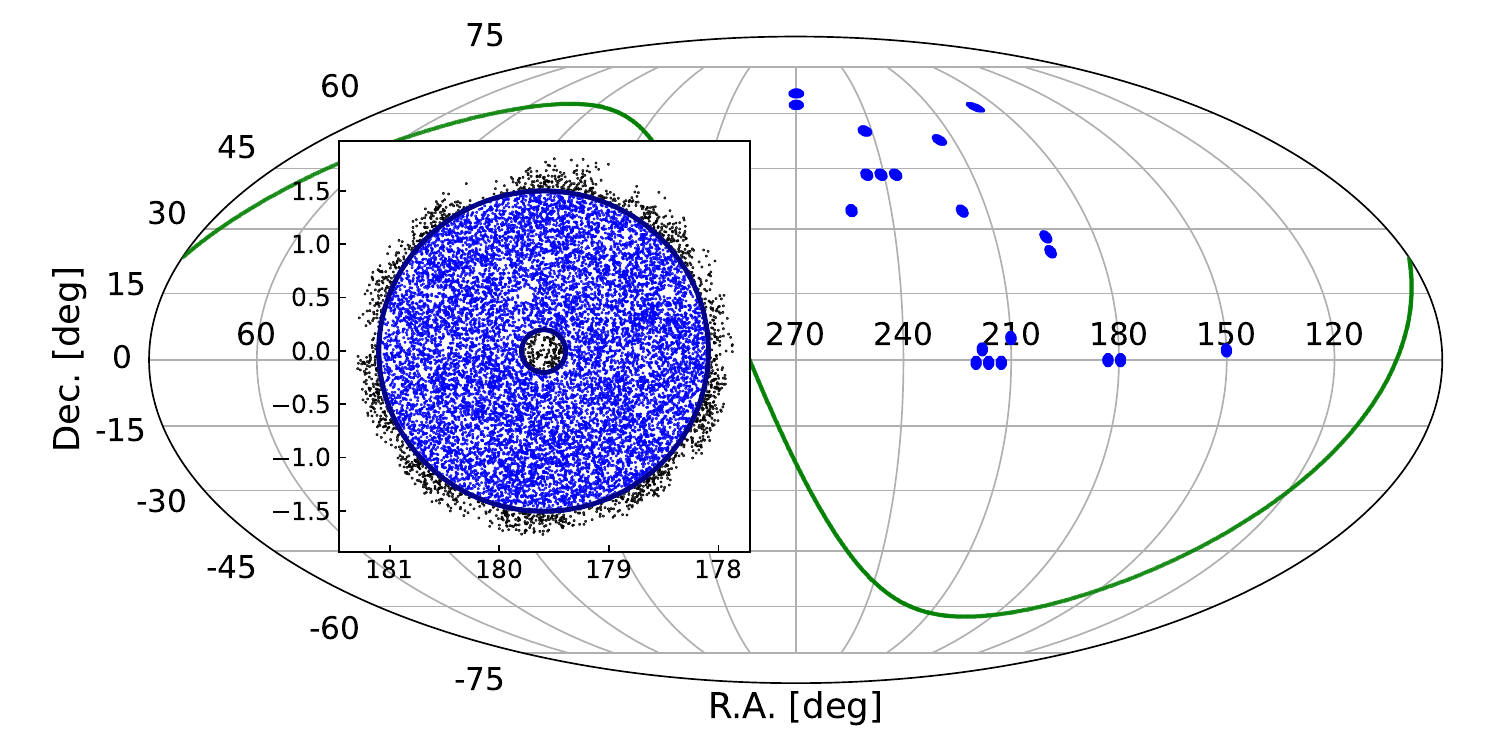}
    \caption{Footprint of the 20 One-Percent Survey rosettes in blue (Mollweide projection). The Galactic plane is shown in green. The inset image shows the angular distribution of spectroscopically confirmed ELGs within one rosette; the galaxies observed in the highly complete annulus are highlighted in blue.}
    \label{fig:footprint}
\end{center}
\end{figure*}

This work utilizes LSS catalogs\footnote{\url{https://data.desi.lbl.gov/public/edr/vac/edr/lss/v2.0/}} from the DESI survey validation program that covers $\sim$140 deg$^2$ of the planned survey area, also known as the One-Percent Survey \citep{desi24_sv}.
The spectroscopic data were processed by the pipeline described in \cite{guy23} and fit by Redrock\footnote{\url{https://github.com/desihub/redrock/}}, the spectral template software, to classify each object and determine redshifts.
These data were made publicly available as part of the DESI Early Data Release \citep{desi24_edr}\footnote{\url{https://data.desi.lbl.gov/doc/releases/edr/}}.
The footprint consists of 20 regions (``rosettes,'' with each being slightly larger than the field of view of the DESI focal plane) and is shown in Figure~\ref{fig:footprint}.
To balance the redshift coverage of the ELGs with other targets during fiber assignment, each ELG is assigned one of two priority designations, low ($\tt{ELG\_LOP}$) or very low ($\tt{ELG\_VLO}$), in the form of targeting bits \citep{myers23}.
Because each rosette in the One-Percent survey received a minimum of 11 observational visits, the ELG sample is highly complete despite having lower priority than other target classes \citep{desi24_sv}.
To perform the clustering measurements, each LSS catalog is accompanied by several ``random'' catalogs that are created to match the data's angular coverage (e.g., regions around bright stars are masked) and redshift distribution but with an otherwise random spatial distribution \citep{ross24}.

Given the distribution of fibers in the DESI focal plane \citep{silber23} and survey tiling strategy \citep{desi24_sv}, targets near the center and outer edge of each rosette receive less fiber coverage.
As in \cite{rocher23}, we define an annulus within each rosette (inset of Figure~\ref{fig:footprint}) that spans 0.2$\degree$ to 1.5$\degree$ radially outward from the rosette center and corresponds to an angular area of $\sim$7 deg$^2$.
Within each annulus, $\sim$95\% of ELG targets had their spectra reliably measured \citep{desi24_sv}, resulting in high completeness.
To construct our overall sample, we use ELGs from the One-Percent LSS catalogs that have either $\tt{ELG\_LOP}$ or $\tt{ELG\_VLO}$ priority designation and are contained within the annuli defined above.

\subsection{ELG Samples} \label{subsec:subsamples}
\begin{figure}
\begin{center}
    \includegraphics[width=0.45\textwidth]{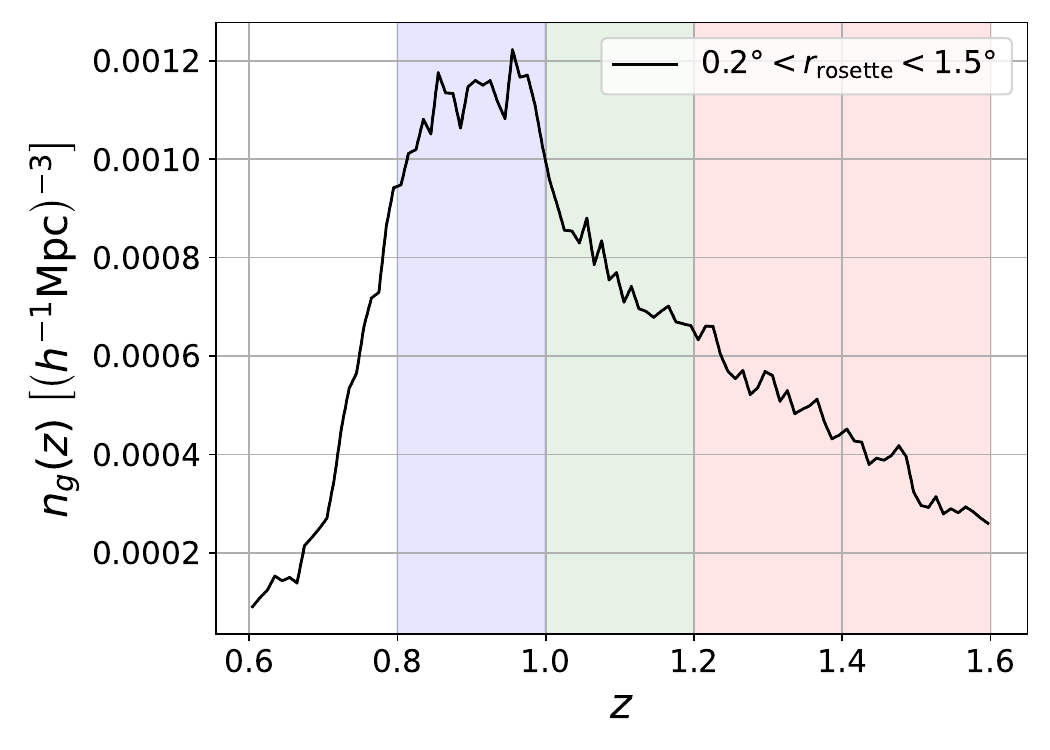}
    \caption{Number density of spectroscopically confirmed ELGs in the DESI One-Percent Survey. Only galaxies within rosette annuli (Figure~\ref{fig:footprint}) are included. Colored regions mark the redshift ranges considered in this work.}
    \label{fig:nz}
\end{center}
\end{figure}

ELG redshifts can be measured quickly and accurately via strong emission lines, and in particular the [OII] doublet at rest-frame wavelengths of 3726 and 3729 $\text{\AA}$ \citep{raichoor23}.
Further, ELGs are abundant at the epoch when the average star formation rate of the Universe peaked, at a redshift of $z \sim 2$ \citep{madau14}, making them especially useful out to $z\sim1.6$ before the [OII] doublet is redshifted out of DESI's wavelength coverage.
The ELG sample targeted by DESI lies between $0.6<z<1.6$, extending to sources 1 magnitude fainter and 10 times denser than that of its predecessor, the Extended Baryon Oscillation Spectroscopic Survey (or eBOSS; \citealt{dawson16, raichoor17}).
This surface density is sufficient to reach the subpercent precision necessary for measurements of the baryon acoustic oscillation feature pursued by DESI \citep{raichoor23}.
The number density of the ELGs in the DESI One-Percent survey as a function of redshift is shown in Figure~\ref{fig:nz}.

The ELG LSS catalogs were combined with value-added catalogs containing stellar mass ($M_*$) and [OII] line flux ($F_{\text{[OII]}}$) estimated by \textsc{FastSpecFit} (Fuji, version 3.2).
In this code, the dust-corrected stellar continua and emission lines are modeled to infer galactic properties using the DESI Legacy photometry and DESI spectroscopy \citep{moustakas23}. Given the model for integrated flux from a galaxy's [OII] doublet, the luminosity is
\begin{equation} \label{eqn:luminosity}
    L_{\rm [OII]} = 4 \pi D_L^2(z) F_{\rm [OII]},
\end{equation}
where $D_L(z)$ is the luminosity distance.

We do not use all ELGs from the LSS catalogs, but rather require those with a $F_{\text{[OII]}}$ signal-to-noise ratio $>3$ to ensure reasonable precision in the [OII] flux measurement (removing an additional $\sim$0.6\% of objects).
The median fractional uncertainty in total [OII] flux is $\sim$6\%. Galaxies were then divided into three redshift ranges: $0.8<z<1.0$, $1.0<z<1.2$, and $1.2<z<1.6$.

Within each redshift range, the physical properties provided by \textsc{FastSpecFit} were compared with those from COSMOS2020, the latest release of the Cosmic Evolution Survey \citep{weaver22, weaver23}.
This comparison revealed that, relative to COSMOS2020, DESI has a measurement scatter of $\sigma_{\log M_*} \sim 0.5$ in stellar mass.
Such a scatter can flatten the $\log L_{\rm [OII]}$-$\log M_*$ distribution of our samples and influence our model constraints.
A detailed discussion is given in Appendix \ref{app:compare_cosmos}.
Uncertainties in the stellar mass estimates and other selection criteria naturally lead to incompleteness in the DESI selection, meaning this work reflects the DESI sample rather than an intrinsic star-forming galaxy population.

\begin{figure*}
\begin{center}
    \includegraphics[width=0.99\textwidth]{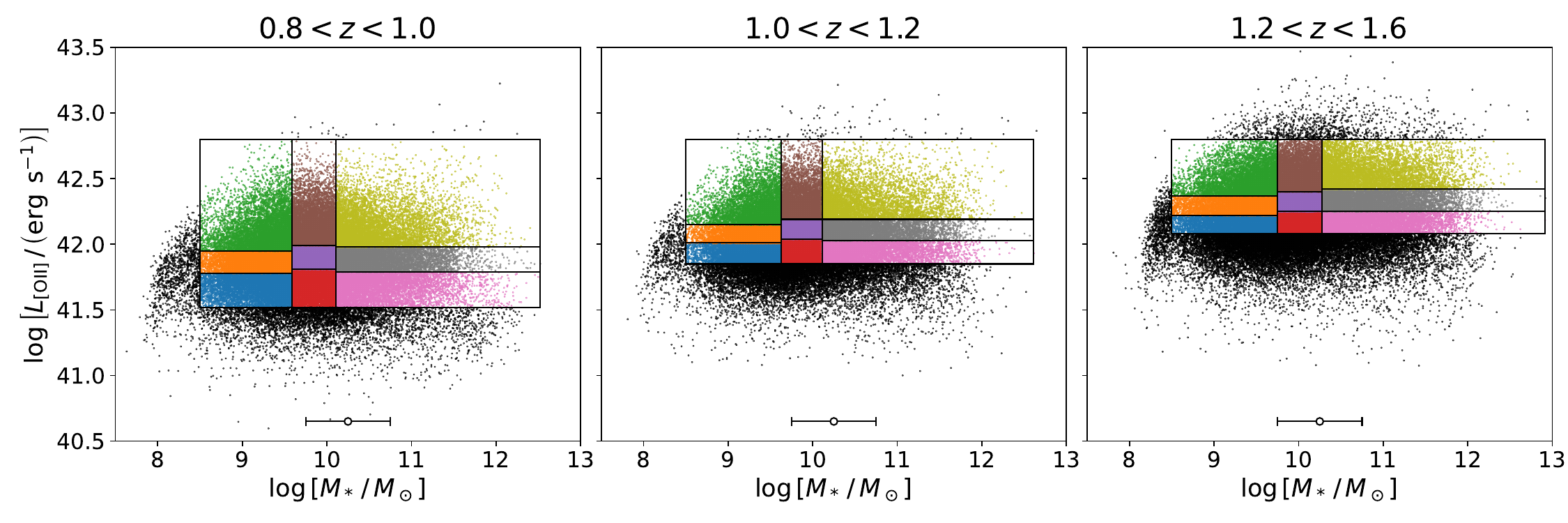}
    \caption{Selection boundaries for the nine samples within each redshift group as a function of $\log L_{\text{[OII]}}$ and $\log M_*$. The horizontal error bars illustrate the 0.5 dex measurement scatter in stellar mass as quantified in Appendix \ref{app:compare_cosmos} with respect to COSMOS2020. Left, middle, and right panels correspond to redshift ranges $0.8<z<1.0$, $1.0<z<1.2$, and $1.2<z<1.6$, respectively.}
    \label{fig:samples}
\end{center}
\end{figure*}

We define nine samples with approximately equal number density in each redshift range according to stellar mass and [OII] luminosity.
The nine samples per redshift are plotted in the \lglg{} plane in Figure~\ref{fig:samples}.
The definitions and properties of each sample are given in Table~\ref{tab:samples}.

\begin{deluxetable*}{lccccccccc}
    \tablecaption{ELG Sample Definitions and Properties\label{tab:samples}}
    \tablehead{
    \colhead{Sample} & \colhead{$\langle z \rangle$} & \colhead{$\log M_{*, \: \rm min}$} & \colhead{$\log M_{*, \: \rm max}$} & \colhead{$\langle \log M_* \rangle$} & \colhead{$\log L_{\rm [OII], \: min}$} & \colhead{$\log L_{\rm [OII], \: max}$} & \colhead{$\langle \log L_{\text{[OII]}} \rangle$} & \colhead{$N_g$} & \colhead{$n_{g}$}
    }
    \startdata
    Z1 Parent & 0.91 & 7.63\tablenotemark{a} & 12.52 & 9.87 & 40.60 & 43.23 & 41.86 & 71,995 & 11.60 \\
    Z1M1L1 & 0.89 & 8.50 & 9.59 & 9.20 & 41.52 & 41.78 & 41.68 & 7448 & 1.18 \\
    Z1M1L2 & 0.91 & 8.50 & 9.59 & 9.22 & 41.78 & 41.95 & 41.86 & 7182 & 1.15 \\
    Z1M1L3 & 0.92 & 8.50 & 9.59 & 9.26 & 41.95 & 42.80 & 42.11 & 7290 & 1.16 \\
    Z1M2L1 & 0.89 & 9.59 & 10.11 & 9.85 & 41.52 & 41.81 & 41.70 & 7561 & 1.22 \\
    Z1M2L2 & 0.91 & 9.59 & 10.11 & 9.84 & 41.81 & 41.99 & 41.90 & 7289 & 1.19 \\
    Z1M2L3 & 0.93 & 9.59 & 10.11 & 9.84 & 41.99 & 42.80 & 42.15 & 7413 & 1.20 \\
    Z1M3L1 & 0.89 & 10.11 & 12.52 & 10.61 & 41.52 & 41.79 & 41.68 & 7166 & 1.16 \\
    Z1M3L2 & 0.91 & 10.11 & 12.52 & 10.58 & 41.79 & 41.98 & 41.88 & 7396 & 1.20 \\
    Z1M3L3 & 0.93 & 10.11 & 12.52 & 10.61 & 41.98 & 42.80 & 42.14 & 7355 & 1.22 \\
    \hline
    Z2 Parent & 1.10 & 7.82 & 12.64 & 9.92 & 41.01 & 43.21 & 42.06 & 59,359 & 7.92 \\
    Z2M1L1 & 1.09 & 8.50 & 9.63 & 9.26 & 41.85 & 42.01 & 41.94 & 5604 & 0.74 \\
    Z2M1L2 & 1.10 & 8.50 & 9.63 & 9.28 & 42.01 & 42.15 & 42.08 & 5261 & 0.70 \\
    Z2M1L3 & 1.11 & 8.50 & 9.63 & 9.31 & 42.15 & 42.80 & 42.29 & 5541 & 0.73 \\
    Z2M2L1 & 1.09 & 9.63 & 10.11 & 9.85 & 41.85 & 42.04 & 41.96 & 5617 & 0.75 \\
    Z2M2L2 & 1.10 & 9.63 & 10.11 & 9.86 & 42.04 & 42.19 & 42.11 & 5127 & 0.69 \\
    Z2M2L3 & 1.11 & 9.63 & 10.11 & 9.87 & 42.19 & 42.80 & 42.33 & 5477 & 0.73 \\
    Z2M3L1 & 1.09 & 10.11 & 12.61 & 10.66 & 41.85 & 42.03 & 41.95 & 5568 & 0.76 \\
    Z2M3L2 & 1.10 & 10.11 & 12.61 & 10.64 & 42.03 & 42.19 & 42.11 & 5387 & 0.72 \\
    Z2M3L3 & 1.11 & 10.11 & 12.61 & 10.65 & 42.19 & 42.80 & 42.34 & 5566 & 0.75 \\
    \hline
    Z3 Parent & 1.38 & 7.82\tablenotemark{a} & 12.99 & 10.04 & 41.08 & 43.81 & 42.26 & 79,572 & 4.50 \\
    Z3M1L1 & 1.35 & 8.50 & 9.75 & 9.33 & 42.08 & 42.22 & 42.15 & 6804 & 0.38 \\
    Z3M1L2 & 1.38 & 8.50 & 9.75 & 9.35 & 42.22 & 42.37 & 42.29 & 7000 & 0.39 \\
    Z3M1L3 & 1.41 & 8.50 & 9.75 & 9.38 & 42.37 & 42.80 & 42.50 & 6740 & 0.38 \\
    Z3M2L1 & 1.35 & 9.75 & 10.28 & 9.99 & 42.08 & 42.25 & 42.17 & 6913 & 0.39 \\
    Z3M2L2 & 1.38 & 9.75 & 10.28 & 10.01 & 42.25 & 42.40 & 42.32 & 6497 & 0.37 \\
    Z3M2L3 & 1.41 & 9.75 & 10.28 & 10.02 & 42.40 & 42.80 & 42.53 & 6727 & 0.38 \\
    Z3M3L1 & 1.37 & 10.28 & 12.91 & 10.85 & 42.08 & 42.25 & 42.17 & 6762 & 0.38 \\
    Z3M3L2 & 1.40 & 10.28 & 12.91 & 10.84 & 42.25 & 42.42 & 42.33 & 7071 & 0.40 \\
    Z3M3L3 & 1.44 & 10.28 & 12.91 & 10.84 & 42.42 & 42.80 & 42.55 & 6595 & 0.39 \\
    \enddata
    \tablecomments{Horizontal lines separate redshift groups: Z1 $\in (0.8, 1.0)$, Z2 $\in (1.0, 1.2)$, Z3 $\in (1.2, 1.6)$. The number of galaxies per sample is listed as $N_g$. The measured number densities, weighted according to Equation~(\ref{eqn:iip}), are listed as $n_g$. Stellar masses, [OII] luminosities, and number densities are in units of solar masses ($M_\odot$), erg per second (erg s$^{-1}$), and $10^{-4}$ ($h^{-1}$Mpc)$^{-3}$, respectively. Parent samples are defined with the annulus and signal-to-noise requirements imposed (Sections \ref{subsec:1percent} and \ref{subsec:subsamples}).}
    \tablenotetext{a}{Excluding fewer than three extreme outliers in the \textsc{FastSpecFit} value-added catalog.}
\end{deluxetable*}

After assigning galaxies to each sample, the number densities $n_g(z)$ no longer match those of the random catalogs constructed for the entire sample.
To match the $n_g(z)$ for each of the nine samples in each redshift bin, objects in the random catalogs were divided into fine redshift bins ($\Delta z=0.01$) and randomly down-sampled. 

\subsection{Aperture Correction} \label{subsec:ap_corr}
Every spectrum measured by DESI undergoes a wavelength-dependent flux correction using previously identified stars as calibration standards \citep{guy23}.
However, because ELGs are spatially extended, the fraction of light captured by DESI's 1.5\arcsec{} diameter fiber differs from that of the standard stars, meaning the amplitude of the calibration vector is not accurate.
An accurate flux calibration for galaxies requires corrections for the seeing at the time of observation and the surface-brightness profile.

We apply a customized aperture correction to the [OII] line flux for ELGs used in this study.
To do so, we must first back out the point-source assumption in the original calibration.
We assume an average seeing of 1\arcsec{} FWHM during spectroscopic observation, for which the DESI fiber will capture $\sim$79\% of a standard star's flux and yield an original scale factor of 1.266.
After factoring this correction out, we apply our customized aperture correction using the DESI Legacy Survey's $r$-band photometry \citep{dey19}, with the central wavelength corresponding to $\sim$3100 $\text{\AA}$ in the rest frame of a galaxy at $z=1$.
We multiply the (now uncorrected) flux by the ratio of total $r$-band flux to the predicted amount within the DESI fiber in 1\arcsec{} FWHM seeing.
The median of this ratio across our samples is $\sim$1.59.
Functionally, our total procedure is
\begin{equation} \label{eqn:rcorrection}
    F_{\rm [OII]}^{\rm corr} = \frac{F_{\rm [OII]}^{\rm biased}}{1.266} \cdot \frac{F_{\mathrm{total,} \: r}}{F_{\mathrm{fiber,} \: r}},
\end{equation}
where $F_{\rm [OII]}^{\rm biased}$ is the \textsc{FastSpecFit} estimate for the integrated flux of the [OII] doublet in the DESI spectrum, $F_{\mathrm{total,} \: r}$ is the total $r$-band flux measured in the Legacy Survey, and $F_{\mathrm{fiber,} \: r}$ is the predicted $r$-band flux within the DESI fiber.
Although our correction assumes the same average seeing for all objects, it removes possible systematic errors introduced by treating all ELGs as if they have the same spatial profile.
The implications of our choice of aperture correction are presented in Appendix \ref{app:aperture}.

\section{Clustering Measurements} \label{sec:clustering}
We first quantify the clustering of galaxies with the 2PCF, which describes the excess probability of finding a galaxy pair with transverse separation $r_p$ and line-of-sight separation $r_\pi$. The 2PCF is measured with respect to a random distribution that shares the same footprint as the galaxy sample. We use the Landy-Szalay estimator to make the measurement \citep{landy93}:
\begin{equation} \label{eqn:landy}
    \xi (r_p, \: r_\pi) = \frac{DD - 2DR + RR}{RR}.
\end{equation}
Here, $DD$ is the fraction of all data-data pairs in a separation bin centered on $\left( r_p, \: r_\pi \right)$. $DR$ and $RR$ are similarly defined but for data-random and random-random pairs. In the transverse direction, evenly spaced logarithmic bins with width of $\Delta \log \left[ r_p / \left( h^{-1} \rm{Mpc} \right) \right] \sim 0.21$ were used. Along the line of sight, evenly spaced linear bins with $\Delta r_\pi = 4$ $h^{-1}$Mpc were used up to a maximum value of $r_{\pi, \: \rm max} = 60$ $h^{-1}$Mpc. 

Due to incompleteness in the DESI fiber assignment, galaxy pairs have unequal probability of being observed.
For example, two galaxies that are only accessible with a single fiber cannot both be observed in a single pass.
To account for such effects, pairwise-inverse-probability (PIP) weights are applied to galaxy pairs when measuring the 2PCF.

PIP weighting was done by following the procedure outlined in \cite{lasker25}.
One-hundred and twenty-eight alternative realizations of the fiber assignment algorithm are performed, in which each target's subpriority (used to split priority ties between targets of the same class) is assigned a new random number.
From these realizations, each galaxy is assigned a \emph{bitwise completeness weight}---a series of 128 bits where each bit indicates whether the galaxy was observed (bit value of 1) or not observed (bit value of 0) in a realization.
The PIP weight of a galaxy pair is one over the fraction of realizations in which both galaxies were observed \citep{bianchi17}:
\begin{equation}
    W_{m,n} = \frac{N_{\rm real}}{1 + \texttt{popcount}\left( W_m^{(b)} \: \& \: W_n^{(b)} \right)},
\end{equation}
where $W_m^{(b)}$ is the bitwise weight of galaxy $m$, $N_{\rm real}$ is the total number of realizations, and \texttt{popcount}($W_m^{(b)} \: \& \: W_n^{(b)}$) counts the number of realizations in which both galaxies were observed. Here, $N_{\rm real}=129$ to account for the 128 alternative realizations and the survey observation itself. An offset of 1 is added in the denominator for the same reason.

PIP weighting is done when counting \textit{galaxy-galaxy} pairs. Following the procedures in \cite{desi24_kp3}, when counting \textit{galaxy-random} or \textit{random-random} pairs, the pair weight becomes the product of the two objects' individual-inverse-probability (IIP) weights. Bitwise weights are converted to IIP weights by
\begin{equation} \label{eqn:iip}
    W_{m} = \frac{N_{\rm real}}{1 + \texttt{popcount}\left( W_m^{(b)} \right)}.
\end{equation}

We then integrated $\xi(r_p, \: r_\pi)$ along the line of sight to calculate the projected 2PCF, which minimizes effects from redshift space distortions (e.g., \citealt{hamilton98}):
\begin{equation} \label{eqn:wp}
    w_p(r_p) = 2 \int_{0}^{r_{\pi, \rm max}} \xi (r_p, r_\pi) \, {\rm d}r_\pi = 2 \sum_i \xi (r_p, r_{\pi,i}) \Delta r_\pi.
\end{equation}
The clustering measurements of each sample were performed using \textsc{Pycorr}\footnote{\url{https://github.com/cosmodesi/pycorr}}, which is DESI's implementation of \textsc{Corrfunc} \citep{sinha19, sinha20}.

For each galaxy sample, the covariance matrix was estimated via 160 jackknife samples, each omitting one-eighth of one survey rosette.
Elements of the matrices were calculated following \cite{zehavi05}:
\begin{equation}\label{eqn:cov}
\text{Cov}(w_{p,i}, w_{p,j}) = \frac{N_{\rm jack}-1}{N_{\rm jack}} \sum_{l=1}^{N_{\rm jack}} (w_{p,i}^l - \overline{w}_{p,i}) (w_{p,j}^l - \overline{w}_{p,j}),
\end{equation}
where $l$ denotes one of the $N_{\rm jack}=160$ jackknife samples and $\overline{w}_{p,i}$ is the $w_p$ measurement at the $i$th $r_p$ bin averaged over all jackknife samples.
We determined only the diagonal and first off-diagonal elements of the covariance matrices because the covariance between measurements at vastly different scales quickly becomes noisy.
All other elements were set to zero, and we assume no covariance between samples.
Analysis using fewer jackknife samples (80 and 40) revealed only percent-level variations in the diagonal elements.

\begin{figure*}
\begin{center}
    \includegraphics[width=0.9\textwidth]{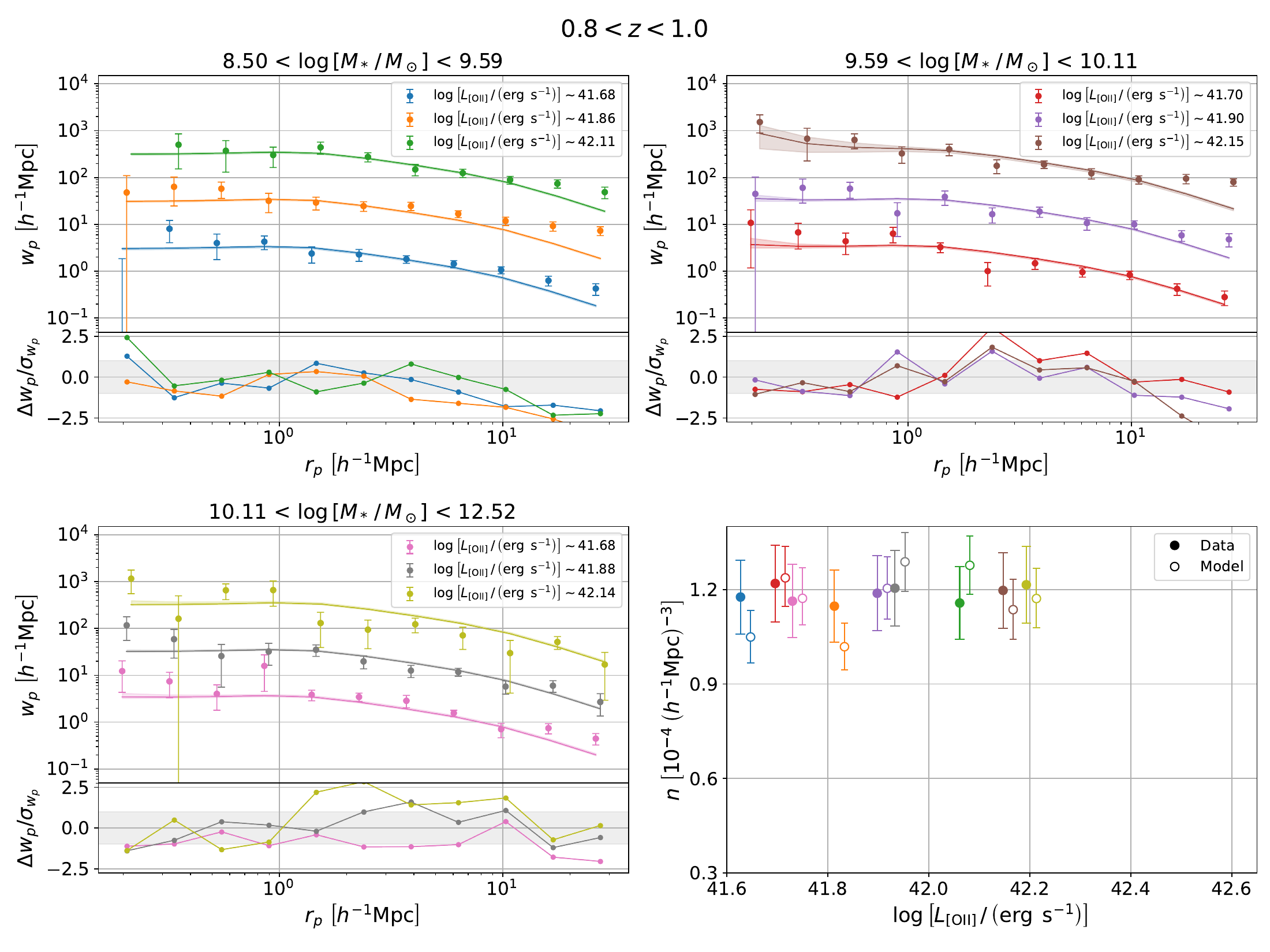}
    \caption{Measurements and fits for the nine $0.8<z<1.0$ samples defined in Figure~\ref{fig:samples} and Table~\ref{tab:samples}. The top-left, top-right, and bottom-left panels show the projected 2PCFs of the low-, intermediate-, and high-stellar-mass samples, respectively. Filled circles mark the measurements. Solid curves and shaded regions mark the median and $1\sigma$ bounds of the posterior model distribution, respectively. For illustrative purposes, measurements are staggered in $r_p$. Vertical offsets of $-1$ dex and $+1$ dex are applied to the lowest and highest [OII] luminosity samples in each panel, respectively. The subpanel below each main panel shows the residual of the model relative to the measurement, normalized by the measurement uncertainty. The shaded region represents the $\pm 1 \sigma$ bounds. \emph{Bottom right:} measurements of the number densities are shown as filled circles, with 10\% uncertainties assumed. Open circles and error bars mark the median and $1\sigma$ bounds of the posterior model distribution, respectively. For illustrative purposes, horizontal offsets are added between the measurement and fit of a given sample and between samples.}
    \label{fig:fit_loz}
\end{center}
\end{figure*}

\begin{figure*}
\begin{center}
    \includegraphics[width=0.9\textwidth]{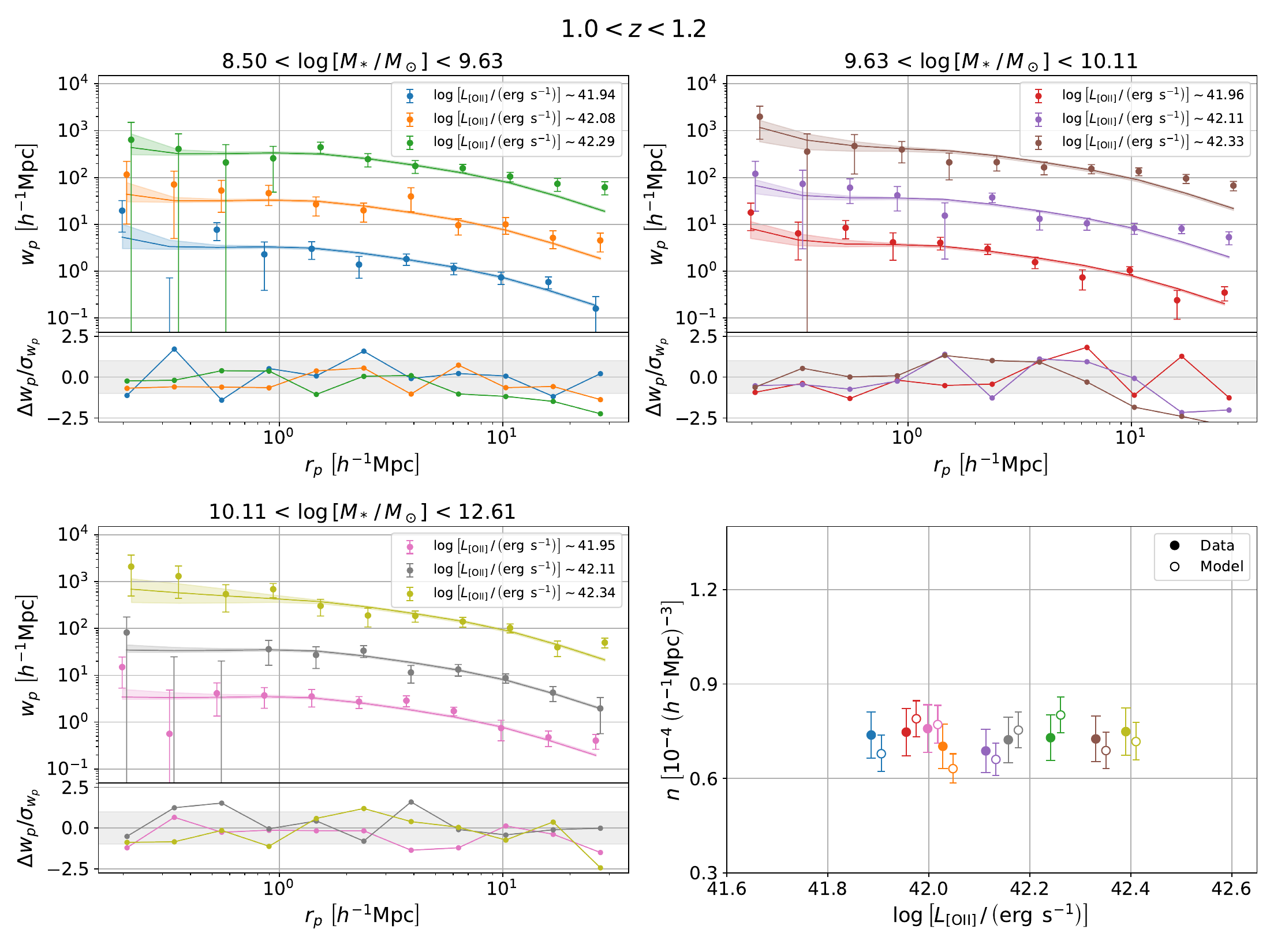}
    \caption{Same as Figure~\ref{fig:fit_loz} but for the $1.0<z<1.2$ intermediate-redshift samples.}
    \label{fig:fit_midz}
\end{center}
\end{figure*}

\begin{figure*}
\begin{center}
    \includegraphics[width=0.9\textwidth]{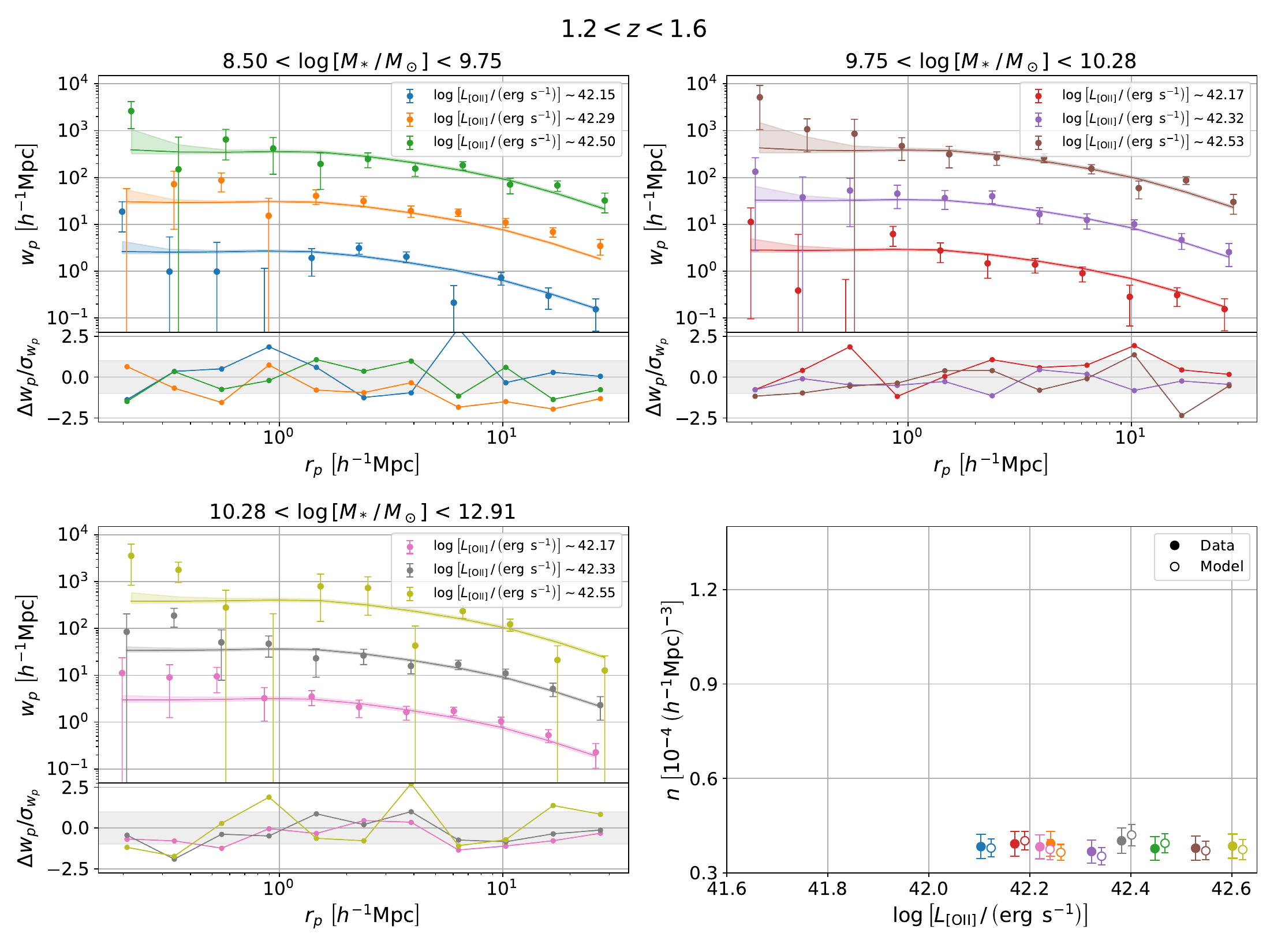}
    \caption{Same as Figure~\ref{fig:fit_loz} but for the $1.2<z<1.6$ high-redshift samples.}
    \label{fig:fit_hiz}
\end{center}
\end{figure*}

Clustering and number density measurements for the low-, intermediate-, and high-redshift samples are shown in Figures~\ref{fig:fit_loz}-\ref{fig:fit_hiz}, respectively.

\section{Modeling the [OII] Luminosity- and Stellar-mass-dependent Clustering} \label{sec:model}
We model the measured [OII] luminosity- and stellar mass-dependent ELG clustering within the halo model framework.
The \lglg{} distribution of our DESI ELG samples is parameterized as a function of halo mass.
Together with analytic descriptions of the halo mass function \citep{jenkins01}, density profile \citep{navarro96}, and bias \citep{sheth01, tinker05}, we predict the ELG clustering as a function of [OII] luminosity and stellar mass.
Here, we present the details of our model, which is applied in Section \ref{sec:results} to our measurements.

\subsection{Occupation Model for Galaxies}
Following the CCMD framework in \cite{xu18} and \cite{clontz22}, we describe the conditional \lglg{} distribution of galaxies occupying halos of mass $M_h$; centrals and satellites are each given a distinct model.
The mean number of central and satellite galaxies occupying halos of a given halo mass are denoted by $\langle N_{\rm cen}(M_h) \rangle$ and $\langle N_{\rm sat}(M_h) \rangle$, respectively.
For ease of reading, we define the logarithmic stellar mass as $x \equiv \log \left[ M_* / M_\odot \right]$ and the logarithmic [OII] luminosity as $y \equiv \log \left[ L_{\rm{[OII]}} \, / \, \left( \rm{erg \;\, s}^{-1} \right) \right]$.
Across the central and satellite models, we have 15 free parameters to describe the conditional \lglg{} distribution of ELGs.
A summary of the model parameters is given in Table~\ref{tab:params}.
We provide a detailed description in what follows.

\subsubsection{Distribution of Central Galaxies} \label{sec:modelcen}
For central galaxies, the distribution at fixed halo mass is parameterized as a 2D Gaussian according to
\begin{equation} \label{eqn:Ncen}
    \frac{\text{d}^2 \langle N_{\rm cen}(M_h) \rangle}{\text{d}x \: \text{d}y} = \frac{A_c}{2 \pi \sigma_x \sigma_y \sqrt{1-\rho^2}} \exp \left[ -\frac{Z^2}{2(1-\rho^2)} \right],
\end{equation}
where
\begin{equation}
Z^2 = \frac{(x - \mu_x)^2}{\sigma_x^2} + \frac{(y - \mu_y)^2}{\sigma_y^2} - \frac{2\rho(x - \mu_x)(y - \mu_y)}{\sigma_x \sigma_y}.
\end{equation}
Here, $A_c$ is the fraction of halos with mass $M_h$ that are occupied by a central ELG, $\mu_x$ and $\mu_y$ are the respective median values of $\log M_*$ and $\log L_{\text{[OII]}}$, $\sigma_x$ and $\sigma_y$ are the respective standard deviations (which describe the combination of intrinsic scatter and measurement uncertainty), and $\rho$ is the correlation coefficient between $\log M_*$ and $\log L_{\text{[OII]}}$.
In this work, we assume $\log M_*$ and $\log L_{\text{[OII]}}$ are uncorrelated at fixed halo mass, meaning $\rho=0$.


The parameters in the above 2D Gaussian equations depend on halo mass following similar descriptions in \cite{xu18} and \cite{clontz22}.
The amplitude of the Gaussian varies as a power law:
\begin{equation} \label{eqn:logAc}
    \log A_c = \log A_{c,p} + \gamma_A (\log M_h - \log M_{h,p}),
\end{equation}
where $M_{h,p}$ is the pivot mass, $A_{c,p}$ is the amplitude at the pivot mass, and $\gamma_A$ is the power-law index.

The median values of stellar mass and [OII] luminosity are parameterized as power laws with exponential cutoffs toward low halo mass.
For the median stellar mass ($M_{*,m}$), this is
\begin{equation}
    M_{*,m} = M_{*,t} \left( \frac{M_h}{M_{tx}} \right)^{\alpha_x} \exp\left[ -\frac{M_{tx}}{M_h} + 1 \right].
\end{equation}
Here, $M_{tx}$ is the transition mass scale, $M_{*,t}$ is the median stellar mass at that scale, and $\alpha_x$ is the power-law index indicating the high-mass slope.
Taking the logarithm ($\mu_x = \log M_{*,m}$) yields
\begin{equation} \label{eqn:mux}
    \mu_x = \mu_{xt} + \alpha_x (\log M_h - \log M_{tx}) + \left( -\frac{M_{tx}}{M_h} + 1 \right) \Big/ \ln 10,
\end{equation}
where $\mu_{xt} \equiv \log M_{*,t}$.
Similarly, the median value of $y$ is
\begin{equation}
    \mu_y = \mu_{yt} + \alpha_y (\log M_h - \log M_{ty}) + \left( -\frac{M_{ty}}{M_h} + 1 \right) \Big/ \ln 10.
\end{equation}
Following \cite{clontz22}, we assume the standard deviations $\sigma_x$ and $\sigma_y$ are constant with respect to halo mass.


\subsubsection{Distribution of Satellite Galaxies} \label{sec:modelsat}
We also describe the \lglg{} distribution of satellites as a function of halo mass.
First considering only stellar mass, the conditional (logarithmic) stellar mass function of satellite galaxies is described by a modified Schechter function in a similar vein to \cite{yang08}:
\begin{equation}
    \frac{\text{d} \langle N_{\rm sat}(M_h) \rangle}{\text{d}x} = \phi_s \left( \frac{M_*}{M_{*,s}} \right)^{\alpha_s + 1} \exp \left[ -\left( \frac{M_*}{M_{*,s}} \right)^2 \right]
\end{equation}
\begin{equation}
    = \phi_s 10^{(\alpha_s + 1)(x-x_s)} \exp \left[ -10^{2(x-x_s)} \right],
\end{equation}
where $\phi_s$ is a normalization, $\alpha_s$ is the slope at low stellar mass, $M_{*,s}$ is the characteristic mass that marks the transition between power-law and exponential behavior, and $x_s \equiv \log{M_{*,s}}$.
Our model has the flexibility to describe a gap between the satellite characteristic stellar mass ($x_s$) and the median central stellar mass ($\mu_x$).
In this work, we fix that gap to zero, making $x_s = \mu_x$.

To incorporate [OII] luminosity, $\log{}L_{\text{[OII]}}$ is described as a Gaussian at fixed $M_*$, with median value $\mu_{y, \: \rm sat}$ and width $\sigma_{y, \: \rm sat}$.
Thus, the \lglg{} distribution of satellites becomes
\begin{equation} \label{eqn:Nsat}
    \begin{split}
    \frac{\text{d}^2 \langle N_{\rm sat}(M_h) \rangle}{\text{d}x \: \text{d}y} & = \frac{1}{\sqrt{2 \pi}\sigma_{y, \: \rm{sat}}} \exp\left[ -\frac{(y - \mu_{y, \: \rm{sat}})^2} {2\sigma_{y, \: \rm{sat}}^2} \right] \\
    & \cdot \phi_s 10^{(\alpha_s + 1)(x-x_s)} \exp\left[ -10^{2(x-x_s)} \right].
    \end{split}
\end{equation}

We parameterize the halo mass dependence of $\phi_s$ in a similar way to Equation~(\ref{eqn:logAc}):
\begin{equation}
    \log \phi_s = \log \phi_{s,p} + \gamma_\phi (\log M_h - \log M_{h,p}).
\end{equation} Given the weak constraining power on $\gamma_\phi$ in our application, we fix $\gamma_\phi$ to unity, making $\phi_s$ linear with $M_h$. 

Motivated by the star formation main sequence \citep{sparre15}, we parameterize $\mu_{y, \: \rm sat}$ to be linearly dependent on $x$:
\begin{equation}
\mu_{y, \: \rm sat} = \mu_{yp, \: \rm sat} + \gamma_{ys}(x - x_p),
\end{equation}
where $\gamma_{ys}$ is the slope and $\mu_{yp, \: \rm sat}$ is the logarithmic [OII] luminosity of satellites at logarithmic stellar mass $x_p=10$.

\begin{deluxetable*}{lcc}
    \tablecaption{Model Parameters and Priors\label{tab:params}}
    \tablehead{
    \colhead{Parameter} & \colhead{Description} & \colhead{Allowed Range}
    }
    \startdata
     & Centrals & \\
    \hline
    $\log M_{h,p}$ & Pivot halo mass for the ELG fraction-halo mass relation. & 11  \\
    $\log A_{c,p}$ & ELG fraction at the pivot halo mass. & $\left[ -2, 0 \right]$  \\
    $\gamma_A$ & Power-law index for the ELG fraction-halo mass relation. & $\left[ -1, 1 \right]$  \\
    $\log M_{tx}$ & Transition halo mass for stellar-halo mass relation. & $\left( -\infty, \infty \right)$ \\
    $\mu_{xt}$ & Median $\log M_*$ at the transition halo mass. & $\left( -\infty, \infty \right)$ \\
    $\alpha_x$ & Power-law index for the stellar-halo mass relation. & $\left[ 0.25, 0.31 \right]$  \\
    $\log \sigma_{x}$ & Logarithmic scatter in $\log M_*$. &  $\left[ -1, 0 \right]$ \\
    $\log M_{ty}$ & Transition halo mass for the [OII] luminosity-halo mass relation. & $\left[ 5, 14 \right]$  \\
    $\mu_{yt}$ & Median $\log L_{\rm [OII]}$ at the transition halo mass. & $\left( -\infty, \infty \right)$ \\
    $\alpha_y$ & Power-law index for the [OII] luminosity-halo mass relation. & $\left( -\infty, \infty \right)$  \\
    $\log \sigma_{y}$ & Logarithmic scatter in $\log L_{\rm [OII]}$. & $\left[ \log 0.005, 0 \right]$  \\
    $\rho$ & \lglg{} correlation coefficient. & 0 \\
    \hline
     & Satellites & \\
    \hline
    $\log \phi_{s,p}$ & Normalization at the pivot halo mass. & $\left( -\infty, \infty \right)$ \\
    $\gamma_{\phi}$ & Power-law index for the normalization-halo mass relation. & 1  \\
    $\alpha_s$ & Slope at low stellar mass. & $\left[ -5, 0 \right]$  \\
    $x_s$ & Characteristic stellar mass of power law-to-exponential transition. & $\mu_x$  \\
    $x_p $ & Pivot stellar mass for the [OII] luminosity-stellar mass relation. & 10 \\
    $\mu_{yp, \: \rm sat}$ & Median $\log L_{\rm [OII]}$ at the pivot stellar mass. & $\left( -\infty, \infty \right)$ \\
    $\gamma_{ys} $ & Power-law index for the [OII] luminosity-stellar mass relation. & $\left( -\infty, \infty \right)$  \\
    $\log \sigma_{y, \: \rm sat}$ & Logarithmic scatter in $\log L_{\rm [OII]}$. & $\left[ \log \sigma_y, 0 \right]$ \\
    \hline
     & Additional Priors & \\
    \hline
    $\log M_{tx} - \log h - \log \left( 1-\alpha_x \right)$ & Peak location of the stellar-halo mass ratio. & $\left[ 11.477, 12.301 \right]$ \\
    $\mu_{x,t} - \log M_{tx} + \log h$ & Peak value of the stellar-halo mass ratio. & $\left[ \log 0.005, \log 0.05 \right]$ \\
    $\mu_{y, \: \rm sat}$ at $\log M_h = 12$ & Median $\log L_{\rm [OII]}$ of satellites. & $\left[ 41, \infty \right)$ \\
    $\frac{\text{d} \langle N_{\rm sat}(M_h) \rangle}{\text{d}x}$ at $\log M_* = 10$ & Conditional stellar mass function of satellites. & $\left[ 10^{-4}, \infty \right)$ \\
    and $\log M_h = 12$ &  & \\
    \enddata
    \tablecomments{Parameters are separated into those for the central and satellite models. When applying our model, we impose several restrictive and additional detailed priors, which are described in Section \ref{sec:application}. Parameters that describe a stellar mass, [OII] luminosity, or halo mass are in units of $M_\odot$, erg s$^{-1}$, or $h^{-1} M_\odot$, respectively.}
\end{deluxetable*}

\subsection{Calculation of Observables and Derived Quantities} \label{sec:calc_dq}
In this work, we adopt the halo mass function in \cite{jenkins01}, defining halos as gravitationally bound regions with a mean density of 200 times the background matter density.
The halo bias is taken from \cite{sheth01} but calibrated according to \cite{tinker05}.
We assume satellite galaxies within a halo follow the dark matter distribution, which is described by a Navarro–Frenk–White profile and depends on the concentration parameter \citep{navarro96},
\begin{equation}
    c(M_h) = \frac{c_0}{1+z} \left( \frac{M_h}{M_{nl}} \right)^\beta,
\end{equation}
with $c_0=11$, $\beta=-0.13$, and $M_{nl} = 3.79 \cross 10^{12}$ $h^{-1} M_\odot$.

For a set of galaxy-halo model parameters, Equations~(\ref{eqn:Ncen}) and (\ref{eqn:Nsat}) can be integrated over each sample's bounds to yield the corresponding $\langle N_{\rm cen}(M_h) \rangle$ and $\langle N_{\rm sat}(M_h) \rangle$.
By combining these galaxy occupation statistics with our halo description, the projected 2PCF of every sample is predicted simultaneously following \cite{zheng04} and \cite{tinker05}, in which the clustering is decomposed into contributions by pairs within single halos and across different halos.
Additionally, the number density of each sample is calculated as
\begin{equation}
    n_g = \int_{0}^{\infty} \langle N_{\rm cen+sat}(M_h) \rangle \frac{\text{d}n}{\text{d}M_h} \text{d}M_h,
\end{equation}
where $\langle N_{\rm cen+sat}(M_h) \rangle \equiv \langle N_{\rm cen}(M_h) \rangle + \langle N_{\rm sat}(M_h) \rangle$ and $\frac{\text{d}n}{\text{d}M_h}$ is the halo mass function.

In our application, we also study several derived quantities of each sample based on the modeling result.
We calculate the large-scale galaxy bias, $b_g$, as
\begin{equation} \label{eqn:bias}
    b_g = \frac{1}{n_g} \int_0^\infty b_h (M_h) \langle N_{\rm cen+sat}(M_h) \rangle \frac{\mathrm{d}n}{\mathrm{d}M_h} \mathrm{d}M_h,
\end{equation}
where $b_h$ is the halo bias.
We also calculate the median halo mass of centrals, $\widetilde{M}_{h, \: \rm cen}$, as
\begin{multline}
    \int_0^{\widetilde{M}_{h,\: \rm cen}} \langle N_{\rm cen}(M_h) \rangle \frac{\text{d}n}{\text{d}M_h} \text{d}M_h = \\
    \frac{1}{2} \int_0^\infty \langle N_{\rm cen}(M_h) \rangle \frac{\text{d}n}{\text{d}M_h} \text{d}M_h.
\end{multline}
The fraction of ELGs that are satellites (known as the satellite fraction) in each sample, $f_{\rm sat}$, is calculated as
\begin{equation}
    f_{\rm sat} = \frac{1}{n_g} \int_0^\infty \langle N_{\rm sat}(M_h) \rangle \frac{\mathrm{d}n}{\mathrm{d}M_h} \mathrm{d}M_h.
\end{equation}
Similar to the case for centrals, the median halo mass of satellites, $\widetilde{M}_{h, \: \rm sat}$, is determined by
\begin{multline} \label{eqn:mh_sat}
    \int_0^{\widetilde{M}_{h, \: \rm sat}} \langle N_{\rm sat}(M_h) \rangle \frac{\text{d}n}{\text{d}M_h} \text{d}M_h = \\
    \frac{1}{2} \int_0^\infty \langle N_{\rm sat}(M_h) \rangle \frac{\text{d}n}{\text{d}M_h} \text{d}M_h.
\end{multline}

For a given redshift, the model simultaneously fits 108 measurements (the 2PCF in 11 $r_p$ bins, plus the number density, for each of nine \lglg{} samples) with 15 free parameters, resulting in 93 degrees of freedom. The $\chi^2$ of the global fit is calculated as

\begin{equation}
    \begin{split}
    \chi^2 = \left( \boldsymbol{w_{p, \: \rm model}} - \boldsymbol{w_{p, \: \rm data}} \right) ^\text{T} \bf{C}_{\boldsymbol{w_p}}^{-1} \left( \boldsymbol{w_{p, \: \rm model}} - \boldsymbol{w_{p, \: \rm data}} \right) \\
    + \left( \boldsymbol{n_{\rm model}} - \boldsymbol{n_{\rm data}} \right) ^\text{T} \bf{C}_{\boldsymbol{n}}^{-1} \left( \boldsymbol{n_{\rm model}} - \boldsymbol{n_{\rm data}} \right).
    \end{split}
\end{equation}

Here, $\boldsymbol{w_{p, \: \rm data}}$ ($\boldsymbol{w_{p, \: \rm model}}$) is the projected 2PCF data (model) vector, $\boldsymbol{n_{\rm data}}$ ($\boldsymbol{n_{\rm model}}$) is the number density data (model) vector, and $\bf{C}_{\boldsymbol{w_p}}$ and $\bf{C}_{\boldsymbol{n}}$ are the respective covariance matrices.
We assume 10\% uncertainty in the measured number density of each sample.
To explore the model parameter space, we employ a Markov Chain Monte Carlo method with a likelihood function proportional to $\exp(-\chi^2 / 2)$.

\section{Results and Discussion} \label{sec:results}
In this section, we present the clustering measurements and modeling results, which are organized by low, intermediate, and high redshift in Figures~\ref{fig:fit_loz}-\ref{fig:fit_hiz}, respectively.

At fixed redshift, the measured number densities are roughly equal by construction, as shown in the bottom-right panels of Figures~\ref{fig:fit_loz}-\ref{fig:fit_hiz}.
The projected 2PCF measurements (other panels of Figures~\ref{fig:fit_loz}-\ref{fig:fit_hiz}) have several notable features.
First, they lack strong trends with respect to stellar mass, [OII] luminosity, or redshift.
This absence is likely a consequence of the measurement scatter in stellar mass ($\sim$0.5 dex as shown in Appendix \ref{app:compare_cosmos}) and a lack of correlation between [OII] luminosity and stellar mass \citep{favole20}.
Second, clustering measurements on the smallest scales ($r_p \lesssim 0.6$ $h^{-1}$Mpc) have large uncertainties.
This reflects the small survey volume and low number densities of our samples, resulting in a dearth of close pairs.
Finally, at fixed redshift and on scales above $r_p \sim 10$ $h^{-1}$Mpc, the projected 2PCFs show various degrees of spurious signal, deviating from the expected nearly constant galaxy bias.
For example, the measurements for low- and intermediate-stellar-mass samples are not parallel to each other.
This may suggest contamination caused by poorly understood imaging systematics.

To interpret the clustering measurements and quantify the dependence on galaxy properties, we apply the model described in Section \ref{sec:model}.
We then present the constraints on our model parameters, discuss the occupation statistics and derived quantities, and compare with previous studies.

\subsection{Application of the Model} \label{sec:application}
Our model describes the conditional \lglg{} distribution of galaxies as a function of halo mass.
The occupation statistics of any sample defined by cuts in stellar mass and [OII] luminosity can then be calculated.
Therefore, at each fixed redshift ($z=0.9$, $1.1$, and $1.4$), we apply this model to all nine samples simultaneously.

Preliminary fitting revealed that our data alone cannot put tight constraints on several parameters.
We impose the priors in Table~\ref{tab:params}; most priors are broad, but some are more restrictive and discussed below.

Motivated by previous studies (e.g., \citealt{behroozi19}), we place restrictive priors on the amplitude and shape of the stellar-to-halo mass relation (SHMR) for central galaxies.
Specifically, the high-mass slope ($\alpha_x$), peak ratio of stellar-to-halo mass, and halo mass at that peak (``Additional Priors'' in Table~\ref{tab:params}) have tight allowed ranges.
The scatters in stellar mass and [OII] luminosity of centrals are given lower limits loosely motivated by the measurement uncertainties.

Given the large uncertainties in the small-scale clustering measurements, we also impose restrictive priors on the satellite parameters (``Additional Priors'' in Table~\ref{tab:params}), which would otherwise be loosely constrained.
Following \cite{clontz22}, we place a lower limit on the conditional stellar mass function of satellites in halos of $\log \left[ M_h \big/ \left( h^{-1} M_{\odot} \right) \right] = 12$, at $\log \left[ M_* \big/ M_{\odot} \right] = 10$.
At the median stellar mass of centrals in halos of $\log \left[ M_h \big/ \left( h^{-1} M_{\odot} \right) \right] = 12$, we also require that the median [OII] luminosity of satellites exceeds $10^{41}$ erg s$^{-1}$.
This ensures that the star formation main sequence of satellites passes through the [OII] luminosities spanned by our samples.

Fits to the low-, intermediate-, and high-redshift measurements are shown in Figures~\ref{fig:fit_loz}-\ref{fig:fit_hiz}, respectively.
Solid curves and shaded regions (top-left, top-right, and bottom-left panels) mark the median and $1\sigma$ bounds of the posterior distribution of predicted 2PCFs.
Open circles and uncertainties (bottom-right panels) mark the same but for the number densities.

Our goodness of fits are affected by the unexpected spurious signal in the projected 2PCFs, particularly at low redshift.
In detail, the model accurately predicts the clustering of high-stellar-mass samples on all scales (with a few exceptions).
However, for most low- and intermediate-stellar-mass samples (especially at lower redshift), the model generally underpredicts the clustering on scales above $r_p \sim 10$ $h^{-1}$Mpc and cannot capture the spurious signal indicated by the measurements.
With 93 degrees of freedom, the best-fit $\chi^2$ values for the low-, intermediate-, and high-redshift fits are 261.7, 99.9, and 97.4, respectively.
The large $\chi^2$ of the low-redshift fit suggests that the uncertainties in the measurements are underestimated due to systematic errors not captured in the covariance estimation.
Averaged over scales above $10$ $h^{-1}$Mpc, the fractional uncertainty of the 2PCF measurements at low redshift are 1.3 (1.6) times smaller than those at intermediate (high) redshift.
These small uncertainties on large scales, in conjunction with stronger systematic discrepancies, yield a greater penalty in $\chi^2$ at low redshift.
In contrast, the measurements at intermediate and high redshift are well fit, indicating the model constraints are appropriately estimated.

For the low-redshift results, rather than scaling the covariance matrix to produce a lower $\chi^2$, we proceed with the caveat that the constraints are not as tight as indicated by the confidence intervals.
As an additional test at low redshift, we recalculate the $\chi^2$ using the above best-fitting model while omitting the two largest-scale 2PCF measurements of each sample.
Now with 75 degrees of freedom, the $\chi^2$ becomes $\sim$95.1, which is within $\sim$1.6$\sigma$ of the expected value.
This test suggests that the low-redshift model results are robust against the spurious signal.

The model favors a clustering amplitude that weakly depends on stellar mass and redshift; only at high redshift, the model shows a noticeable dependency on [OII] luminosity.
The predicted number densities are consistent with the measurement, and the posterior widths are comparable to the measurement uncertainties.

\subsection{Constraints on Model Parameters} \label{subsec:constraints}
\begin{figure*}
\begin{center}
    \includegraphics[width=0.99\textwidth]{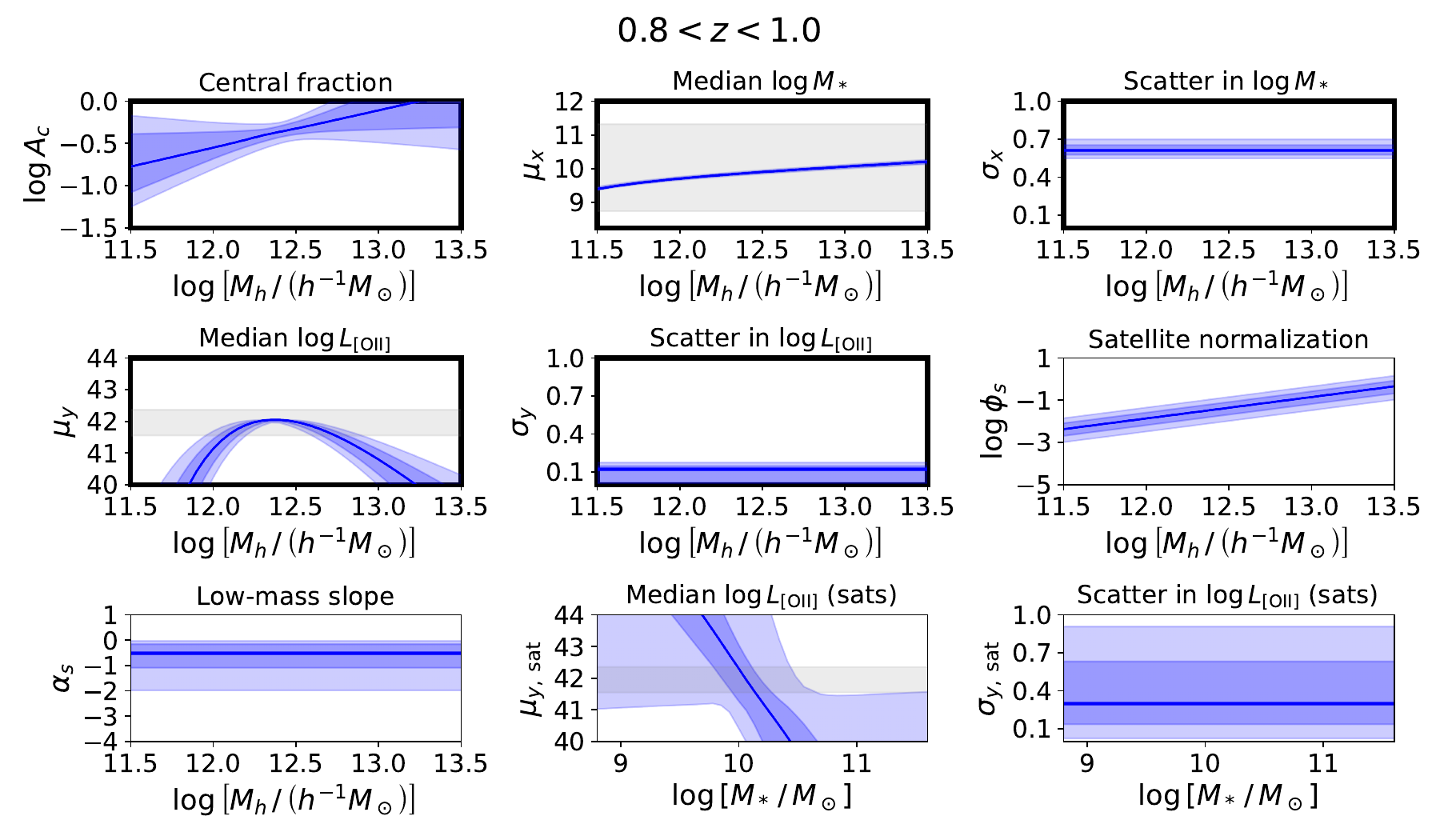}
    \caption{Model constraints from the low-redshift ($0.8<z<1.0$) samples. From left to right, the model relations are (\emph{top row}) the occupation fraction of central ELGs, the median $\log M_*$ of centrals, the scatter in $\log M_*$ of centrals, (\emph{middle row}) the median $\log L_{\rm [OII]}$ of centrals, the scatter in $\log L_{\rm [OII]}$ of centrals, the normalization for the conditional stellar mass function of satellites, (\emph{bottom row}) the low-mass slope for the conditional stellar mass function of satellites, the median $\log L_{\rm [OII]}$ of satellites, and the scatter in $\log L_{\rm [OII]}$ of satellites. Solid lines indicate the median solution. Dark and light shaded regions mark the $1\sigma$ and $2\sigma$ bounds, respectively. Panels for $\mu_x$, $\mu_y$, and $\mu_{y, \: \rm sat}$ also include gray shaded regions that mark the central 95\% range spanned by the sample data. Panels for the constraints on central galaxies have thicker outlines.}
    \label{fig:params_loz}
\end{center}
\end{figure*}

\begin{figure*}
\begin{center}
    \includegraphics[width=0.99\textwidth]{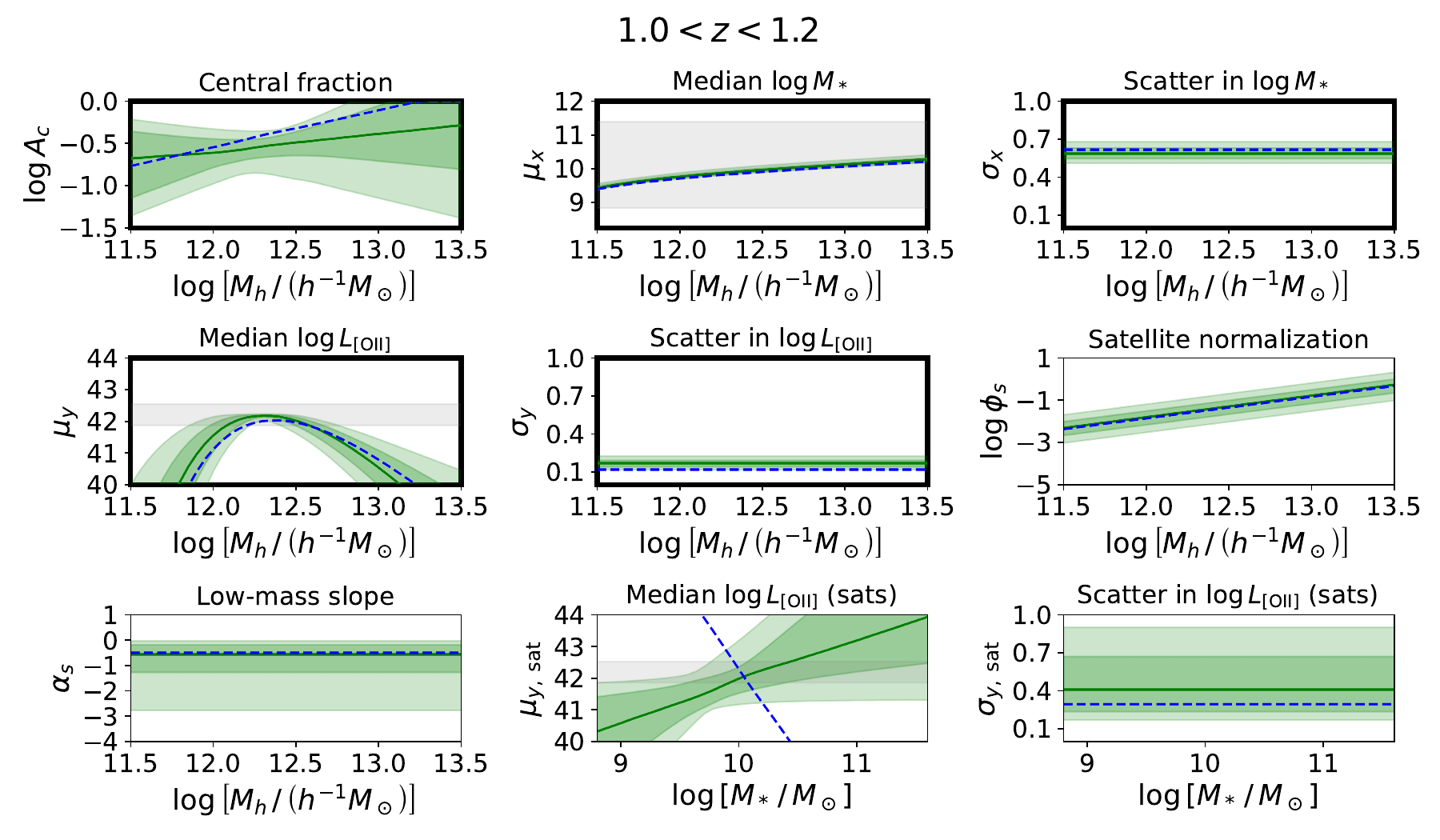}
    \caption{Same as Figure~\ref{fig:params_loz} but for the $1.0<z<1.2$ intermediate-redshift samples. For comparison, the median results from the low-redshift samples (Figure~\ref{fig:params_loz}) are shown as blue dashed lines.}
    \label{fig:params_midz}
\end{center}
\end{figure*}

\begin{figure*}
\begin{center}
    \includegraphics[width=0.99\textwidth]{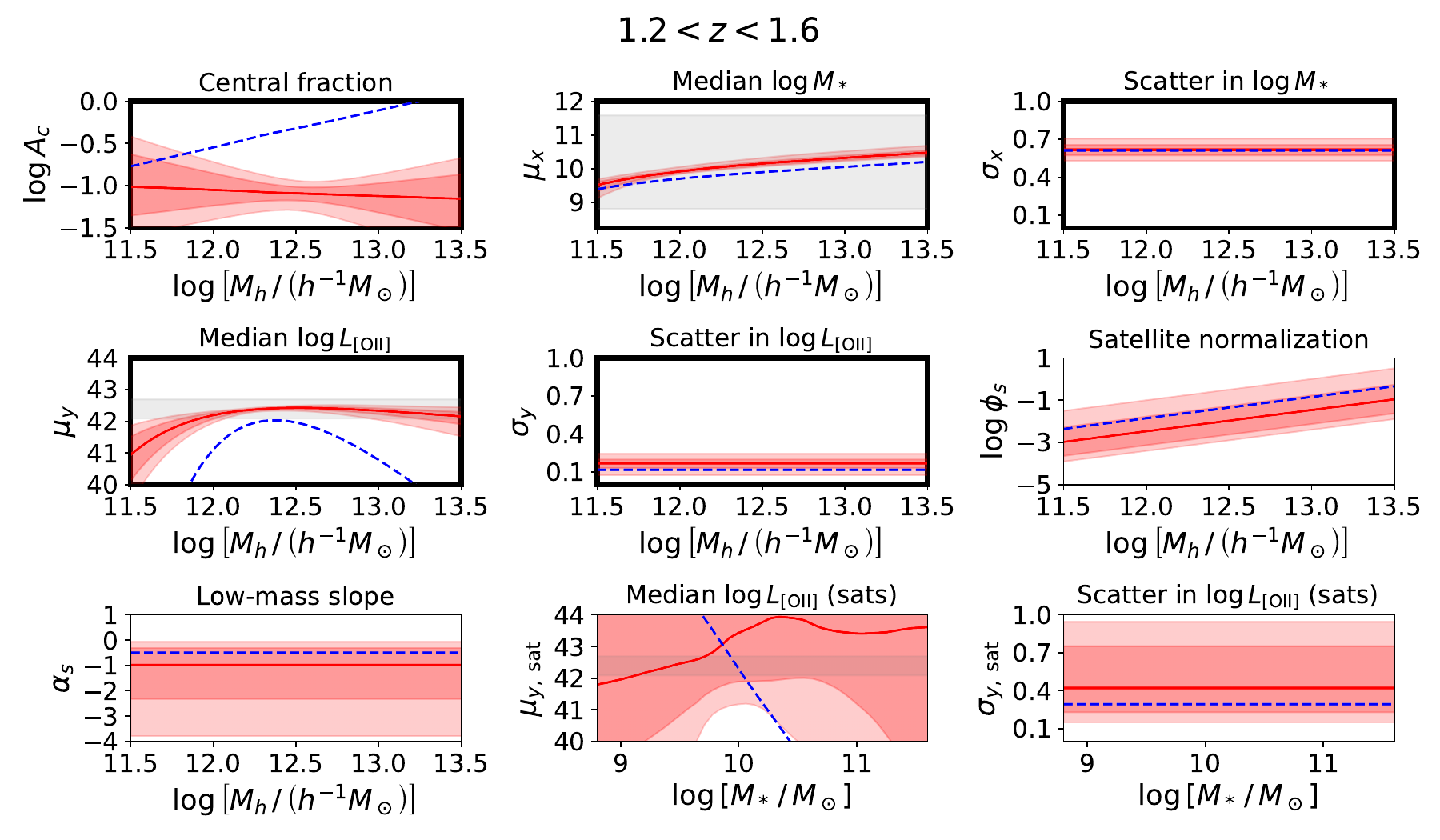}
    \caption{Same as Figure~\ref{fig:params_midz} but for the $1.2<z<1.6$ high-redshift samples.}
    \label{fig:params_hiz}
\end{center}
\end{figure*}

With constraints placed on the model parameters, we present overall relations to halo mass or stellar mass (according to Section \ref{sec:model}) in Figures~\ref{fig:params_loz}-\ref{fig:params_hiz} for the low-, intermediate-, and high-redshift samples, respectively.
Constraints on \textit{all} model parameters are given in Appendix \ref{app:corner}.
We first discuss the low-redshift central galaxy constraints in Figure~\ref{fig:params_loz} (thickly outlined panels).

The fraction of halos hosting an ELG as their central galaxy ($A_c$) increases with halo mass, reaching a regime in which a non-negligible fraction of high-mass halos host a central ELG.
The ELG central fraction is expected to decrease with halo mass as quiescent galaxies generally occupy high-mass halos.
However, for reasons described below, $A_c$ couples with other parameters to impose a strict halo mass range in our samples, which changes the original meaning of $A_c$.

The median stellar mass of centrals ($\mu_x$) increases with halo mass and has a shape in general agreement with literature (see Section \ref{subsec:comparison}).
The scatter in $\log M_*$ of centrals ($\sigma_x$) is larger than the intrinsic value often reported.
\cite{wechsler18} compile results from hydrodynamical simulations, semi-analytic models, and empirical models, all of which predict an intrinsic scatter of $\sigma_x \sim 0.1-0.3$ for central galaxies.
However, our model captures the intrinsic \textit{and} measurement scatter, producing a constraint of $\sigma_x \sim 0.6$ that is consistent with an intrinsic scatter combined with measurement scatter of $\sim$0.5 as quantified in Appendix \ref{app:compare_cosmos}.

The median [OII] luminosity of centrals ($\mu_y$) increases at low halo mass, flattens within the [OII] luminosity range spanned by our samples (gray band), and decreases at high halo mass.
We performed several tests to verify that the model prefers this shape over a flat or monotonically increasing function, which might be expected if [OII] luminosity traces the star formation main sequence.
The parabolic behavior is driven by the interplay between $\mu_y$ and $A_c$.
With $\mu_y$ steeply decreasing toward low and high halo mass where $A_c$ remains nonzero, galaxies in our sample are placed into halos that span a small range of mass.
Regardless of how the halo mass selection is performed, the end result remains the same: The model prefers a narrow range of halo masses to produce central ELGs in our samples.
The scatter in the \lgl{} of centrals ($\sigma_y$) is $\sim$0.1 dex, consistent with the small measurement uncertainty.

Moving to the satellite constraints (thinly outlined panels), the amplitude of the satellite distribution ($\phi_s$) increases with halo mass but is small, meaning the fraction of galaxies in our samples parameterized as satellites is low.
This constraint reflects the relatively flat small-scale clustering in our 2PCF measurements and indicates a weak one-halo term.
Due to the low satellite fraction, all other satellite parameters ($\alpha_s$, $\mu_{y, \: \rm sat}$, and $\sigma_{y, \: \rm sat}$) are poorly constrained.

To highlight potential redshift evolution in our constraints, the median low-redshift results are overlaid in Figures~\ref{fig:params_midz} and \ref{fig:params_hiz} as dashed blue lines.
The constraints at intermediate redshift (Figure~\ref{fig:params_midz}) are consistent with those at low redshift.

At high redshift (Figure~\ref{fig:params_hiz}), $A_c$ and $\mu_y$ behave differently.
In particular, $A_c$ is small and decreases with halo mass, while $\mu_y$ remains flat at high halo mass instead of turning over.
The change in these two relations highlights their degeneracy as described above.
At high redshift, $A_c$ is used to exclude high-mass halos, allowing $\mu_y$ to remain flat and within the [OII] luminosity range of our samples.
At each redshift, either combination of behaviors can be imposed with little penalty in $\chi^2$.
This degeneracy is further illustrated in Section \ref{subsec:hod}, where consequences in the occupation statistics are discussed.

\subsection{Implications on the Occupation Statistics} \label{subsec:hod}
\begin{figure*}
\begin{center}
    \includegraphics[width=0.99\textwidth]{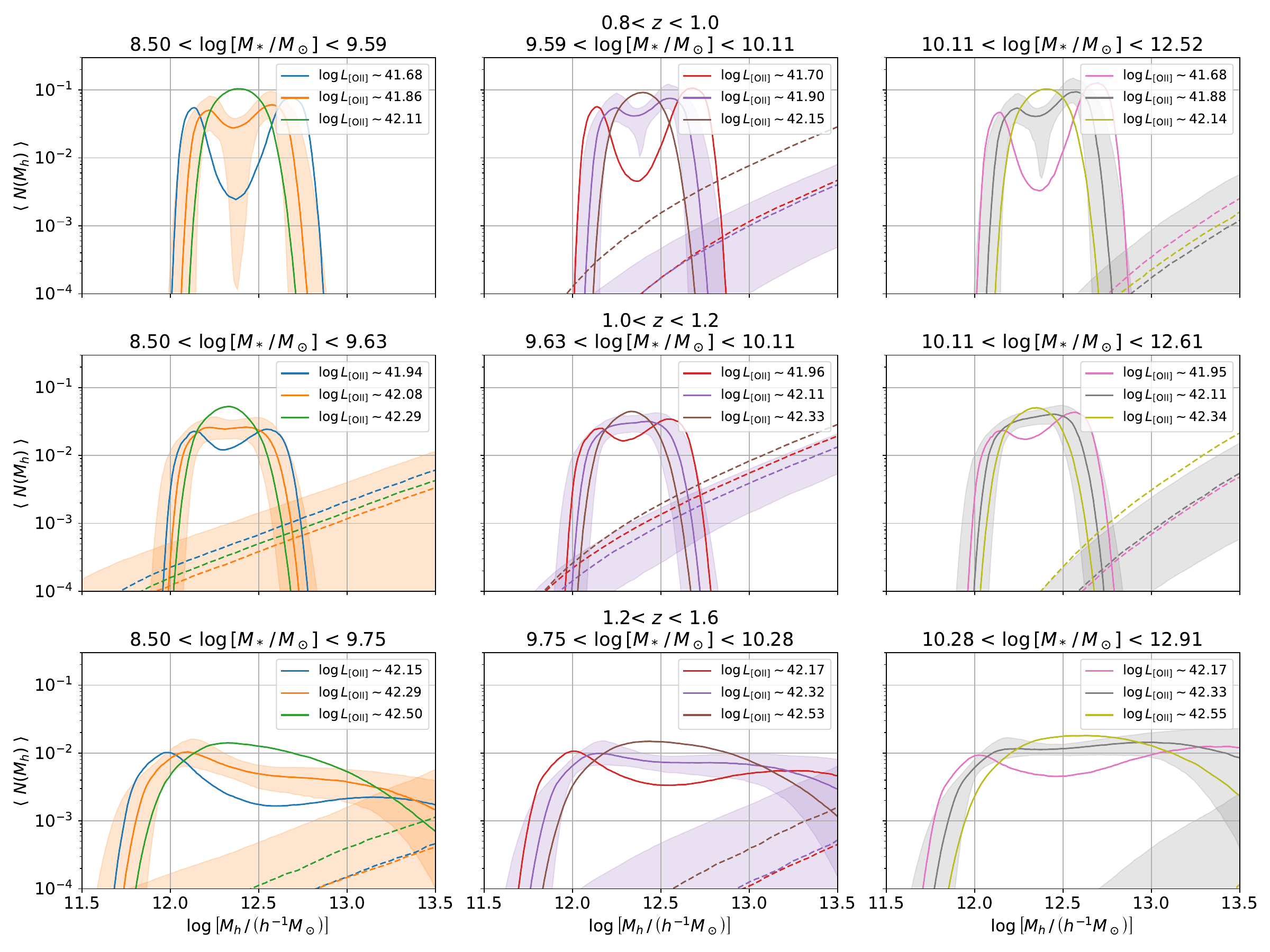}
    \caption{Mean occupation functions of each sample. Each row contains one redshift group. Solid lines and shaded regions show the median and $1\sigma$ bounds of the mean occupation functions for centrals, respectively. Dashed lines mark the same but for satellites. For clarity, uncertainties are included only for the middle [OII] luminosity samples. Error bars for other samples within a panel are roughly the same magnitude. [OII] luminosities are given in erg s$^{-1}$.}
    \label{fig:hod}
\end{center}
\end{figure*}

With constraints placed on the model, Equations~(\ref{eqn:Ncen}) and (\ref{eqn:Nsat}) are integrated over each sample's bounds to compute the mean occupation functions of centrals and satellites, respectively.
We show the occupation statistics of each sample as functions of halo mass in Figure~\ref{fig:hod}.
Our samples contain a low fraction of satellites, as discussed in Section \ref{subsec:constraints} and reinforced by the low satellite occupation functions (dashed lines) in Figure~\ref{fig:hod}.

All central occupation functions (solid lines) at low and intermediate redshift span a narrow range of halo masses, $\sim$$10^{12}$ $h^{-1} M_\odot$ to $10^{12.7}$ $h^{-1} M_\odot$.
At high redshift, the central occupation functions have tails extending to higher halo mass but preserve the same median value by also adding low-mass halos.

At all redshifts, a double-peak behavior that softens with increasing [OII] luminosity is observed, which is a direct consequence of the turnover in the mean [OII] luminosity-halo mass relation ($\mu_y$; see Section \ref{subsec:constraints}).
Given the [OII] luminosity range of the lowest [OII] luminosity samples, $\mu_y$ crosses the range at two distinct halo mass scales, causing a doubly peaked occupation function.
The nonzero values of $\sigma_y$ serve to smooth the distribution.
A similar result occurs for the intermediate [OII] luminosity samples, but the two halo mass scales are closer together and the peaks in the occupation functions are less separated.
In the case of high [OII] luminosity samples, $\mu_y$ approaches but never exceeds the samples' [OII] luminosity range, creating a singly peaked occupation function with a maximum at the turnover scale.
This behavior is less evident at high redshift because $\mu_y$ flattens at higher halo mass while $A_c$ suppresses the number of galaxies in high-mass halos.

\begin{figure*}
\begin{center}
    \includegraphics[width=0.8\textwidth]{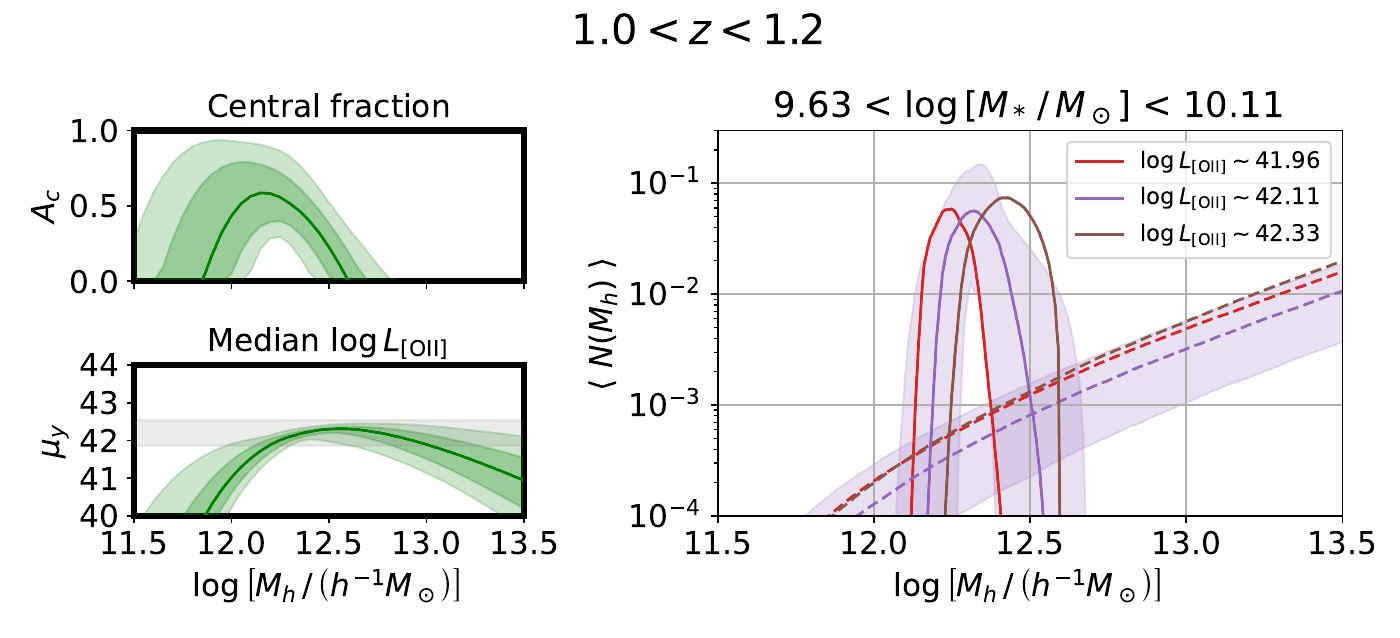}
    \caption{Results from an alternative parameterization of the $A_c$-$M_h$ relation. For illustrative purposes, only constraints for the intermediate-redshift samples are shown. Left panels show the occupation fraction of central ELGs (top) and median $\log L_{\rm [OII]}$ of centrals (bottom), formatted as in Figure~\ref{fig:params_loz}. The right panel shows the mean occupation functions of the intermediate-stellar-mass samples, formatted as in Figure~\ref{fig:hod}. All other constraints are largely unchanged.}
    \label{fig:alt_Ac}
\end{center}
\end{figure*}

As described earlier, the model uses the freedom in $\mu_y$ to apply a halo mass selection that cannot be captured by $A_c$ due to its monotonicity.
To push all selection effects into $A_c$, alternative parameterizations with varying flexibility were tested.
One example is shown in Figure~\ref{fig:alt_Ac}, where $A_c$ varies with halo mass similar to $\mu_y$, providing a means to asymmetrically decrease at both low and high halo mass.
The model has minimal preference toward the original parameterization in Figure~\ref{fig:params_midz} ($\chi^2 \sim 99.9$ with 93 degrees of freedom) over that in Figure~\ref{fig:alt_Ac} ($\chi^2 \sim 101.0$ with 92 degrees of freedom), indicating that the double-peak feature in the occupation function is insignificant.
In all cases, the data push the model toward a scenario in which DESI preferentially observes ELGs within a narrow range of halo masses centered on $\sim$$10^{12.4}$ $h^{-1}M_\odot$.
This preference may be a consequence of the $grz$ selection used to define DESI ELG targets \citep{raichoor23}.
By imposing a faint cut on the $g$-band fiber magnitude of $g_{\rm fib} < 24.1$, faint ELGs in low-mass halos lack enough UV luminosity to be targeted by DESI.
Similarly, by requiring blue colors in $g-r$ and $r-z$, quiescent galaxies that dominate high-mass halos do not enter the selection.

Additional property dependence in the mean occupation functions is evident at high redshift.
The mean halo mass increases with [OII] luminosity, indicated by the changing shapes of the occupation functions.
There are no trends with stellar mass, which is likely due to the large measurement scatter captured by the model.
Theoretically, more precise estimates of stellar mass could reveal trends that are currently unobservable.
Inspection of the derived quantities in Section \ref{subsec:dqs} further supports the presence of galaxy property dependence in the high-redshift samples.

\subsection{Implications on the Derived Quantities} \label{subsec:dqs}
\begin{figure*}
\begin{center}
    \includegraphics[width=0.99\textwidth]{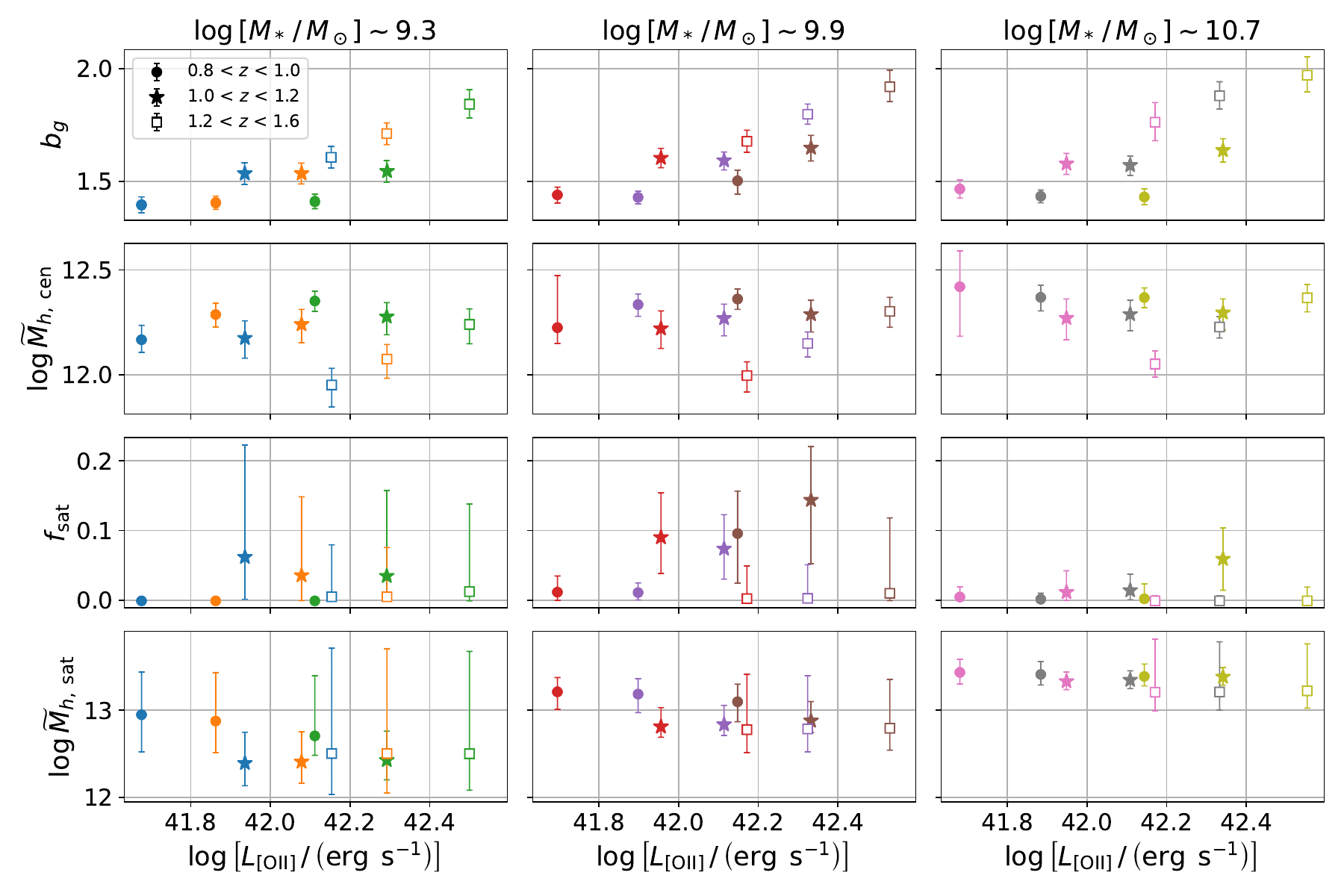}
    \caption{Derived quantities of each sample. From top to bottom, the rows are the galaxy bias, the median halo mass of centrals, the satellite fraction, and the median halo mass of satellites. Halo masses are in units of $h^{-1} M_\odot$. Each panel contains all luminosity and redshift samples at fixed stellar mass. Data points are colored according to the binning in stellar mass and [OII] luminosity in Figure~\ref{fig:samples}. From low to high, redshifts are marked as filled circles, filled stars, and open squares.}
    \label{fig:dq_samples}
\end{center}
\end{figure*}

Given the occupation statistics, several derived quantities are computed according to Equations~(\ref{eqn:bias})-(\ref{eqn:mh_sat}) and shown in Figure~\ref{fig:dq_samples} for each sample.

Galaxy bias (first row in Figure~\ref{fig:dq_samples}) is tightly constrained and generally increases with stellar mass, especially at high redshift.
At lower redshifts, the correlation with stellar mass exists but is weaker.
Trends with [OII] luminosity are more redshift dependent.
At low and intermediate redshift, galaxy bias is nearly constant with [OII] luminosity.
However, at high redshift, galaxy bias shows an increase with [OII] luminosity, with the highest-stellar-mass sample having lower significance.
To quantify this trend, we perform a linear fit to the galaxy bias as a function of (logarithmic) [OII] luminosity at fixed stellar mass and redshift (i.e., three data points are used in each linear fit).
All cases at low and intermediate redshift have best-fitting slopes with an absolute value less than 0.04 dex$^{-1}$.
In contrast, the best-fitting slopes at high redshift are much larger: $\left( 0.72 \pm 0.15\right) $ dex$^{-1}$, $\left( 0.73 \pm 0.21 \right)$ dex$^{-1}$, and $\left( 0.43 \pm 0.28 \right)$ dex$^{-1}$ for the low-, intermediate-, and high-stellar-mass samples, respectively.
The apparent [OII] luminosity dependence at only high redshift might instead be a consequence of our redshift binning; above $\log \left[ L_{\rm{[OII]}} \, / \, \left( \rm{erg \;\, s}^{-1} \right) \right] \sim 42.2$, galaxy bias also increases at lower redshifts, but this trend is most evident at high redshift due to the overall higher [OII] luminosities in that bin. 
Finally, we find that at fixed stellar mass and [OII] luminosity, galaxy bias increases with redshift.

At all redshifts, the median halo mass of centrals (second row in Figure~\ref{fig:dq_samples}) shows an increase with stellar mass that softens at high [OII] luminosity.
Central halo mass also increases with [OII] luminosity, but the prevalence is redshift dependent.
At low and intermediate redshift, the trend flattens at high stellar mass, while the high-redshift case remains strong.
At fixed stellar mass and [OII] luminosity, central halo mass decreases with redshift.

All satellite fractions (third row in Figure~\ref{fig:dq_samples}) are consistent with zero, with those of some samples being loosely constrained.
As a consequence, the median halo mass of satellites (fourth row in Figure~\ref{fig:dq_samples}) is poorly constrained for many samples, especially at low stellar mass.
With few galaxies classified as satellites, any trends observed in the median halo mass of satellites depend on the model parameterization itself and have limited meaning.

\subsection{Comparison With Other Studies} \label{subsec:comparison}
We compare our work to other studies of ELGs from the DESI One-Percent Survey, including those combining with other tracers and those inspecting ELGs alone.
We note that the entire DESI ELG sample is often considered in other works, but we divide the ELGs into bins according to galaxy property, which limits exact comparisons.

\cite{prada25} use a modified SHAM approach to place DESI ELGs in halos and produce mock lightcones.
Their mean halo occupation function for centrals resembles a Gaussian that peaks at $\sim$10$^{11.7}$ $h^{-1} M_\odot$, and that for the satellites increases like a power law, approaching unity at $\sim$$10^{14.6}$ $h^{-1}M_\odot$.
Both shapes are consistent with those predicted by semi-analytical models in \cite{gonzalez-perez20}.

The central occupation functions produced by our model (Figure~\ref{fig:hod}) also span a small range of halo mass.
As discussed in Section \ref{subsec:hod}, the double-peak shape observed in several of our occupation functions is insignificant, and the single-peak profile from alternative parameterizations (Figure~\ref{fig:alt_Ac}) is similar to that in \cite{prada25}.
The more meaningful quantity to compare is the overall halo mass scale.
Our central ELGs consistently occupy higher-mass halos than those in \cite{prada25}.
We find that this discrepancy is largely due to differences in the satellite fraction.
Our model classifies nearly all galaxies as centrals; in contrast, the satellite fraction in \cite{prada25} is $\sim$13\%.
With a larger satellite fraction, the central occupation function must shift to lower halo masses to compensate for the higher bias induced by the satellites.
Using the halo bias in \cite{tinker05}, we estimate that with the total galaxy bias fixed, decreasing the satellite fraction in \cite{prada25} to 0\% will shift the peak of their central occupation function to $\sim$10$^{12.2}$ $h^{-1} M_\odot$, which agrees more with our halo mass scale.
Additionally, our sample cuts remove $\sim$10\% of ELGs with the lowest [OII] luminosities and stellar masses, which might increase the halo mass scale in our model.

\cite{prada25} also report, for the entire ELG data set, galaxy biases that are lower than those of our samples defined by galaxy property (Figure~\ref{fig:dq_samples}).
As a test, we performed the clustering measurement on the combination of our nine samples (at fixed redshift) and found a lower bias that is more consistent with \cite{prada25}.
For example, the projected 2PCF amplitude of the combined low-redshift samples at $r_p \sim 30$ $h^{-1}$Mpc is, on average, $\sim$1.9 times lower than those of the individual samples.
The excess clustering of our ELG samples on such large scales indicates the likely presence of systematic errors that requires further investigation (e.g., by studying the cross correlation between samples).
On smaller scales, the clustering is less affected by such systematic errors, and we can still obtain meaningful constraints on the model.

Regarding satellite fractions, this study is not unique in finding low values for ELGs.
\cite{yu24} apply a SHAM method to ELGs in the DESI One-Percent Survey and report satellite fractions below $6\%$ and consistent with zero at the $2.5\sigma$ level.
Beyond DESI, HOD modeling of galaxies with the highest [OII] luminosity in the VIMOS Public Extragalactic Redshift Survey \citep{scodeggio18} has revealed satellite fractions of $\sim$$5 \% \pm 2 \%$ \citep{gao22}.
While our result is in agreement with a satellite fraction as low as zero, many samples have loose constraints that are also consistent with up to $\sim$$10-20\%$ as found in other studies (e.g., \citealt{lin23, prada25}).
For a proper comparison with any of these studies, the modeling details and sample definitions must be closely examined.

\cite{gao23} model the central SHMR as a double power law and constrain the relation by fitting the stellar-mass-dependent clustering of ELGs and luminous red galaxies.
The stellar masses used in their work are derived by fitting the DESI broadband photometry with \textsc{CIGALE} \citep{boquien19}, which produces stellar masses with an overall offset and smaller dispersion than those used here.
In addition, the extent of the offset depends on stellar mass at the high-stellar-mass end.
Despite parameterizing the SHMR with a shape similar to \cite{gao23}, we find that these differences in stellar mass make it difficult to directly compare constraints.
Therefore, our SHMR should be interpreted specifically for DESI ELGs with stellar masses derived via \textsc{FastSpecFit}.

Galaxy conformity, defined as the correlation of properties between neighboring galaxies such as centrals and satellites \citep{weinmann06}, and its role in the small-scale clustering of the DESI ELGs have been explored in several studies.
\cite{yuan25} present a physical argument for conformity, in which ELGs have preferentially undergone merger activity, resulting in perturbed morphology and boosted star formation.
Such tidal-force-driven star formation increases the likelihood that satellites near ELGs are also ELGs, enhancing the clustering on scales of a few hundred kiloparsecs.
\cite{rocher23} find that in order to accurately fit the strong small-scale clustering of DESI ELGs below $r_p \sim 0.1$ $h^{-1}$Mpc, conformity is required in the model.
Similarly, \cite{gao24} report an improved fit to DESI ELG clustering below $r_p \sim 0.3$ $h^{-1}$Mpc when including conformity.
Because we divide the data by galaxy property, the clustering on such small scales is not well measured, limiting our ability to probe conformity.
Larger data sets from DESI will enable us to constrain conformity and its possible dependence on galaxy property.
Our constraints and interpretation of satellite fractions may also be sensitive to the inclusion of conformity (or lack thereof) as discussed in other studies \citep{rocher23, ortega-martinez24, yuan25}.

\section{Conclusions}\label{sec:conclusions}
In this paper, we measure and model the clustering of ELGs from the DESI One-Percent Survey and its dependence on stellar mass and [OII] luminosity.
In each of three redshift ranges ($0.8 < z < 1.0$, $1.0 < z < 1.2$, $1.2 < z < 1.6$), the projected 2PCFs and number densities of nine $\log L_{\rm [OII]} - \log M_*$ samples are measured.
To fit the measurements of all samples at fixed redshift simultaneously, we apply a modified CCMD framework \citep{xu18, clontz22} that describes the conditional $\log L_{\rm [OII]}-\log M_*$ distribution as a function of halo mass.
We obtain constraints on physically motivated model parameters, mean halo occupation functions, and derived quantities.

The model demonstrates flexibility in fitting the measurements across stellar mass and [OII] luminosity.
Several features in the projected 2PCFs influence our fits.
The 2PCF measurements plateau on small scales and have large uncertainties.
The model responds with satellite fractions that are low or loosely constrained, making constraints on only the central galaxy parameters meaningful.
On scales above $r_p \sim 10$ $h^{-1}$Mpc, varying degrees of spurious signal are observed in the 2PCFs of different samples, suggesting the presence of systematic errors in the imaging data and resulting selection of targets.
This effect is most evident at low redshift, which leads to a $\chi^2$ value of 261.7 (93 degrees of freedom).
Nonetheless, the measurements at intermediate and high redshift are well fit.
Finally, the 2PCFs show only weak dependence on galaxy properties, which is largely reflected in the model constraints.

We summarize our findings with three key takeaways.
First, ELGs in the DESI One-Percent Survey reside in halos that span a narrow range in mass.
In our work, that range is $\sim$$10^{12}$ $h^{-1} M_\odot-10^{12.7}$ $h^{-1} M_\odot$ and accompanied by low satellite fractions.
When compared to other studies, the exact halo mass scale depends on the satellite fraction estimate, which largely explains discrepancies with, e.g., \cite{prada25}.
The existence of such a selection on halo mass is consistent with the occupation functions in other works \citep{gonzalez-perez20, prada25} and can be understood through the DESI target selection \citep{raichoor23}.
The upper limit on $g$-band fiber magnitude removes faint ELGs that reside in low-mass halos, while cuts in $g-r$ and $r-z$ colors exclude quiescent galaxies that dominate high-mass halos.
In our model, this halo mass selection is reproduced by distorting the ELG central fraction ($A_c$) and median [OII] luminosity-halo mass relation ($\mu_y$) to remove galaxies from halos outside the narrow mass range.

Second, we find that the quality of the galaxy property estimates influences our analysis.
The stellar masses have large measurement scatter, which is likely caused by limits in the imaging data and can hide potential dependence in the clustering.
The [OII] luminosities also have a rather weak effect on the clustering, challenging the expectation that [OII] emission is a proxy for star formation rate.
As a consequence, the parameter describing the stellar-to-halo mass scatter ($\sigma_x$) is inflated and those describing the [OII] luminosity-halo mass relation are unexpectedly used to impose a selection on halo mass.
Therefore, our results should be interpreted specifically for DESI ELGs and properties derived by \textsc{FastSpecFit}; extensions of this analysis to general star-forming galaxies should account for these findings.

Third, the model can capture property dependence despite the measurement scatter described above.
Some [OII] luminosity dependence and hints of stellar mass dependence are observed in the occupation functions and derived quantities, especially at high redshift.
Galaxy bias increases with stellar mass and [OII] luminosity at high redshift, while the trends soften or flatten at lower redshift.
The median halo mass of centrals follows a similar trend as a consequence of the low satellite fractions found for all samples.
These results indicate that our model can inform the galaxy-halo connection of DESI ELGs.

Now that DESI has completed 3 yr of observation, we are considering several modifications to our analysis for the larger data sample.
Given the lack of strong property dependence found in this work, other bases beyond \lglg{} will be explored.
In particular, directly observed properties such as magnitudes, colors, and morphology are of interest because they inform the DESI ELG target selection algorithm \citep{raichoor23} and do not rely on modeling of the spectrophotometry.
Joint sampling according to these properties will be examined in the upcoming DESI Year 3 sample to identify the basis with the most influence on the clustering measurements.
Sample-dependent systematic weights should also be assigned to account for variations in the imaging quality.
Such corrections may address the spurious signal that is observed in the projected 2PCF measurements at low redshift and stellar mass.
The chosen basis may also inform changes to our model parameterization.

Regardless of basis, larger data releases by DESI will enable improved sample statistics and stricter quality selection.
Utilizing the entirety of a larger sample will allow us to improve the small-scale measurement and sample more finely in the galaxy properties. 
Alternatively, precision in the galaxy property estimates can be prioritized by using overlaps with deeper and more diverse photometry such as Data Release 3 of the Hyper Suprime-Cam Subaru Strategic Survey (or HSC-SS; \citealt{aihara22}).

Finally, the model presented here relies on analytic descriptions of halo clustering \citep{navarro96, jenkins01, sheth01, tinker05}.
With a larger data sample, we will turn to a more accurate and efficient simulation-based method to model the clustering measurement \citep{zheng16, xu18}.

Within the context of DESI, ELGs will constitute the largest fraction of the full 5 yr sample.
To optimize cosmological constraints from the ELG sample, we should understand the physical properties of these tracers and the impact on their selection, clustering, and relation to the underlying matter distribution.
Our constraints and future studies are catered specifically to DESI and can inform mock galaxy catalogs and galaxy evolution models for future analysis.
Beyond DESI, continued investigation of the ELG-halo connection will remain critical as ELGs are leveraged to study the $z \gtrsim 1$ Universe in future spectroscopic surveys, such as those with Euclid \citep{laureijs11} and the Nancy Grace Roman Space Telescope \citep{green12}.

\section*{Acknowledgments}\label{sec:acknowledgements}
This material is based upon work supported by the U.S. Department of Energy (DOE), Office of Science, Office of High-Energy Physics, under Contract No. DE–AC02–05CH11231, and by the National Energy Research Scientific Computing Center, a DOE Office of Science User Facility, under the same contract. Additional support for DESI was provided by the U.S. National Science Foundation (NSF), Division of Astronomical Sciences, under Contract No. AST-0950945 to the NSF’s National Optical-Infrared Astronomy Research Laboratory; the Science and Technology Facilities Council of the United Kingdom; the Gordon and Betty Moore Foundation; the Heising-Simons Foundation; the French Alternative Energies and Atomic Energy Commission (CEA); the National Council of Humanities, Science and Technology of Mexico (CONAHCYT); the Ministry of Science, Innovation and Universities of Spain (MICIU/AEI/10.13039/501100011033); and by the DESI Member Institutions (\url{https://www.desi.lbl.gov/collaborating-institutions}). Any opinions, findings, and conclusions or recommendations expressed in this material are those of the author(s) and do not necessarily reflect the views of the U.S. National Science Foundation, the U.S. Department of Energy, or any of the listed funding agencies.

The authors are honored to be permitted to conduct scientific research on I'oligam Du'ag (Kitt Peak), a mountain with particular significance to the Tohono O’odham Nation.

The COSMOS2020 data used here are based on observations collected at the European Southern Observatory under
ESO program ID 179.A-2005 and on data products produced by CALET and
the Cambridge Astronomy Survey Unit on behalf of the UltraVISTA consortium.

The support and resources from the Center for High Performance Computing at the University of Utah are gratefully acknowledged.

The work of T.H. and K.D. was supported in part by the U.S. Department of Energy, Office of Science, Office of High Energy Physics, under Award No. DESC0009959. Z.Z. is supported by NSF grant AST-2007499.

\section*{Data Availability}
All data shown in figures are publicly available on Zenodo (doi:\href{https://doi.org/10.5281/zenodo.15467956}{10.5281/zenodo.15467956}).

\bibliographystyle{aasjournal}
\bibliography{references}

\appendix
\section{Comparison with COSMOS2020} \label{app:compare_cosmos}
COSMOS2020 \citep{weaver22, weaver23} provides photometry for $\sim$1.7 million sources within 2 deg$^2$ and is accompanied by a value-added catalog (\textsc{Farmer}) that includes [OII] luminosities and stellar masses predicted by the photometric redshift code \textsc{EAZY} \citep{brammer08} and spectral templates derived from the flexible stellar population synthesis models in \cite{conroy09} and \cite{conroy10}.
One rosette from the DESI One-Percent Survey overlaps the COSMOS2020 footprint, providing a useful comparison between the \textsc{FastSpecFit} and \textsc{EAZY} parameter estimates.

\begin{figure}
\centering
\begin{subfigure}[t]{0.325\textwidth}
    \centering
    \includegraphics[width=0.95\textwidth]{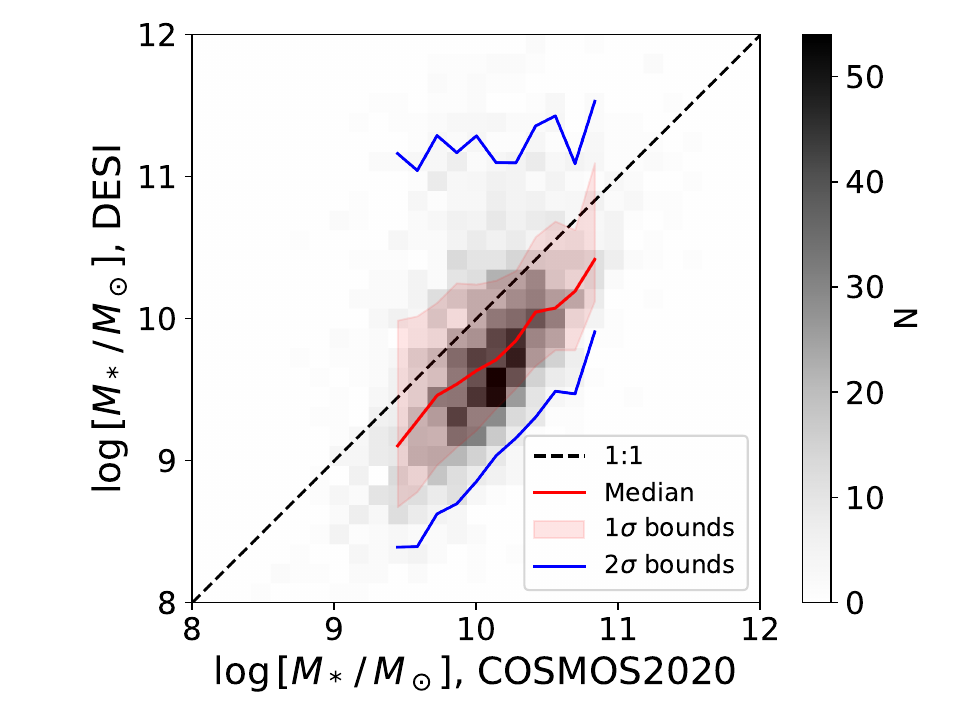}
    \label{fig:mass_desi_cosmos}
\end{subfigure}
\begin{subfigure}[t]{0.325\textwidth}
    \centering
    \includegraphics[width=0.95\textwidth]{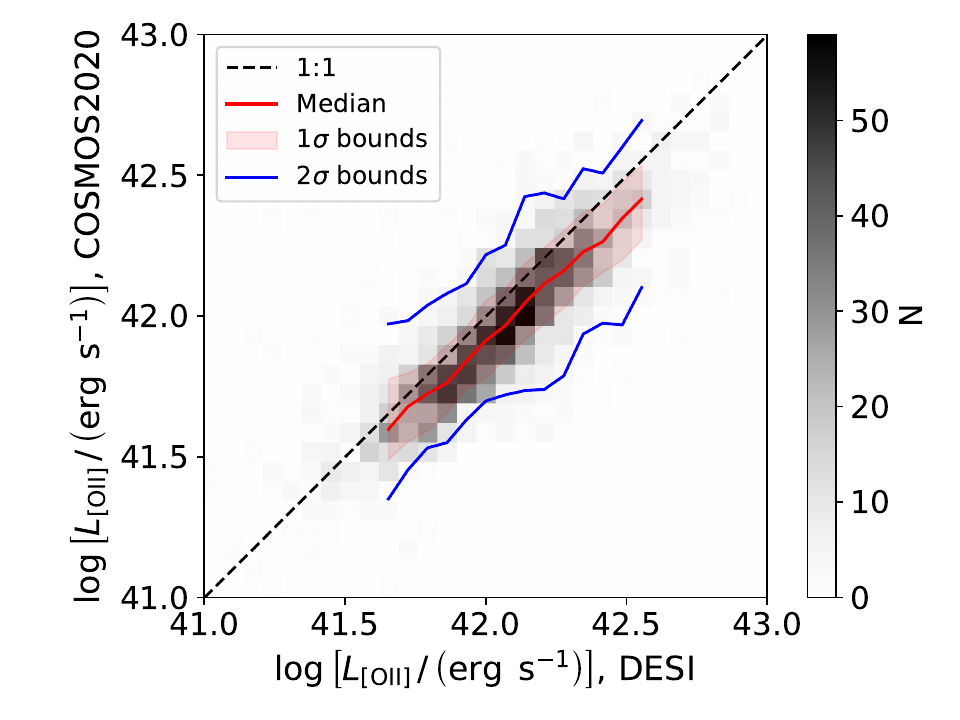}
    \label{fig:lum_desi_cosmos}
\end{subfigure}
\begin{subfigure}[t]{0.325\textwidth}
    \centering
    \includegraphics[width=0.95\textwidth]{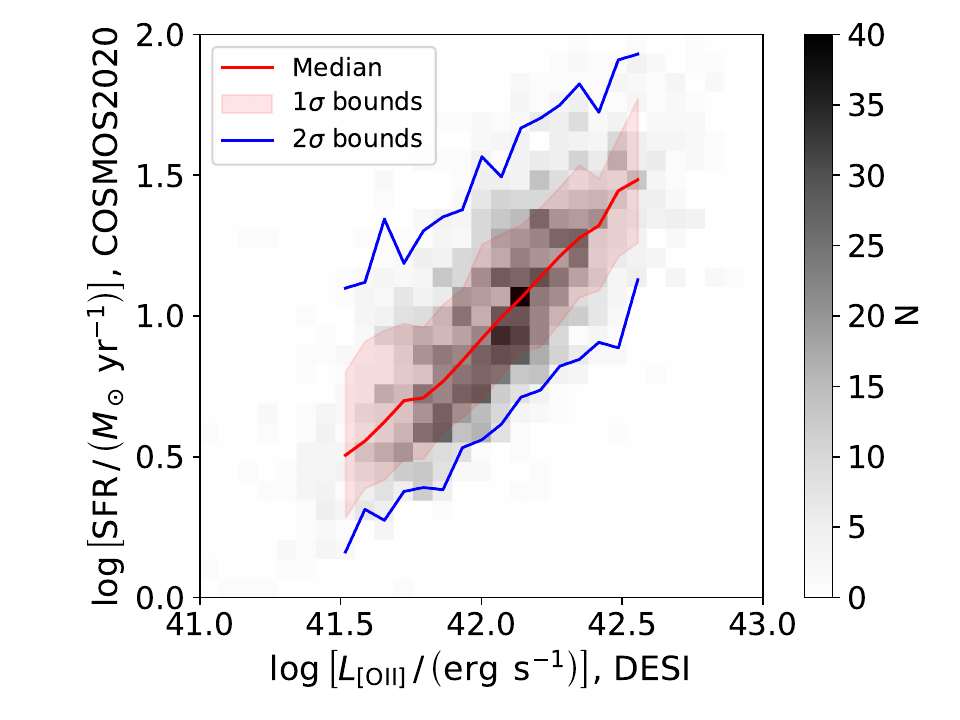}
    \label{fig:sfr_desi_cosmos}
\end{subfigure}
\caption{Comparison of ELG properties reported by DESI (via \textsc{FastSpecFit}) and COSMOS2020 (via \textsc{EAZY}). \emph{Left:} DESI stellar masses have a notable offset and scatter with respect to COSMOS2020. The 2$\sigma$ bounds highlight that the DESI stellar mass distribution is skewed, especially at low stellar masses. \emph{Middle:} COSMOS2020-inferred [OII] luminosities as a function of DESI-measured [OII] luminosities. \emph{Right:} star formation rate estimates from COSMOS2020 as a function of DESI [OII] luminosity.}
\label{fig:desixcosmos}
\end{figure}

We show ELG stellar mass and [OII] luminosity estimates from DESI and COSMOS2020 in Figure~\ref{fig:desixcosmos} (left and middle panels, respectively).
DESI stellar masses have been converted to match COSMOS2020's value of $h$, and both assume a Chabrier initial mass function \citep{chabrier03}.
As a function of COSMOS2020 stellar mass, DESI stellar masses exhibit an offset of roughly -0.37 dex and a scatter of $\sigma_{\log M_*} \sim 0.5$ dex.
This DESI scatter exceeds the statistical uncertainty reported by COSMOS2020 ($\sim$0.04 dex on average) despite the COSMOS2020 estimates relying on photometric redshifts.
The DESI stellar mass estimates are also skewed toward higher values at fixed COSMOS2020 stellar mass, and the extent of this tail is greatest at low stellar mass.
With respect to DESI [OII] luminosities (applying the aperture correction in Section \ref{subsec:ap_corr}), the COSMOS2020 [OII] luminosities have an offset of -0.1 dex and a scatter of $\sigma_{\log L_{\rm [OII]}} \sim 0.13$ dex.

We also plot the COSMOS2020 star formation rate estimates against the DESI [OII] luminosities in Figure 13 (right panel) to inspect the validity of [OII] luminosity as a tracer for star formation rate.
Here, star formation rate is positively correlated with DESI [OII] luminosity but has a scatter of $\sigma_{\log \rm SFR} \sim0.23$ dex and skewness, suggesting that [OII] luminosity is a limited tracer of star formation.
Other studies have also found limitations in [OII] luminosity as such a tracer (e.g., \citealt{moustakas06}).

The subtleties in the DESI ELG properties characterized above lead to consequences in our work, two of which we describe below.

\begin{figure*}
\begin{center}
    \includegraphics[width=0.9\textwidth]{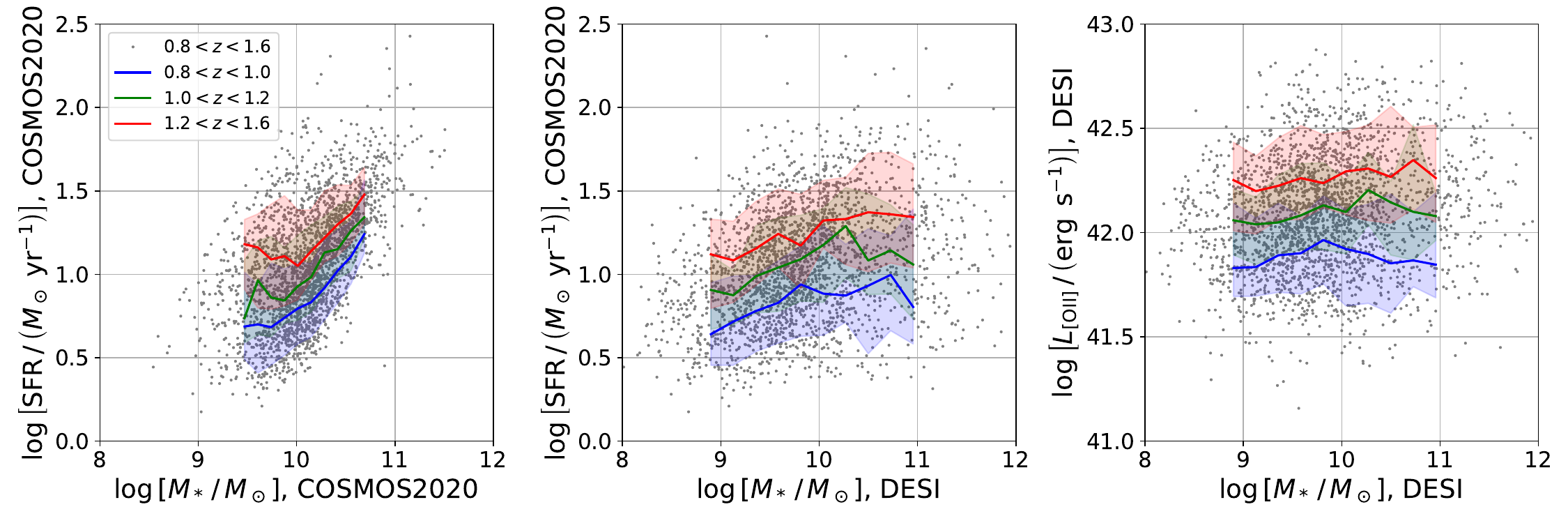}
    \caption{The star formation main sequence with different parameter estimates. Solid lines and shaded regions show the median and 1$\sigma$ bounds of each relation binned by redshift, respectively. \textit{Left:} COSMOS2020 estimates of star formation rate and stellar mass are positively correlated. \textit{Middle:} this relation becomes shallower when DESI stellar mass estimates are instead used. \textit{Right:} further using DESI [OII] luminosity as a proxy for star formation rate yields a flat \lglg{} distribution.}
    \label{fig:ml_comparison}
\end{center}
\end{figure*}

First, the scatter in the DESI stellar mass estimates and limited validity of [OII] luminosity as an indicator of star formation rate suppress evidence of the star formation main sequence (e.g., \citealt{daddi07, noeske07, sparre15}) in the DESI \lglg{} plane.
This effect is demonstrated in Figure~\ref{fig:ml_comparison}, where an originally clear star formation main sequence (left panel) becomes shallower when using the noisier DESI stellar mass estimates (middle) and ultimately flattens in the DESI \lglg{} plane (right).

Second, the offset, scatter, and skewness of the DESI stellar mass distribution can affect our HOD modeling result.
For example, the large stellar mass estimate uncertainty inflates our model constraints on the stellar mass scatter at fixed halo mass ($\sigma_x$ as defined in Section \ref{sec:modelcen}) beyond the intrinsic value of $\sigma_{\log M_*} \sim0.2$ typically inferred from HOD modeling (e.g., \citealt{wechsler18}).
Additionally, the skewness of the stellar mass estimates can alter the shape of the inferred SHMR.

\section{Aperture Correction} \label{app:aperture}
As described in Section \ref{subsec:ap_corr}, many ELGs are spatially extended and require a customized aperture correction to recover missing [OII] flux that is not captured by the DESI fiber. 
Here, the choice of correction and its effects on the [OII] luminosity distribution are explored.

In this work, the measured flux is corrected using Equation~(\ref{eqn:rcorrection}) and $r$-band photometry from the DESI Legacy Imaging Surveys \citep{dey19}.
An alternative correction can be derived by combining the equivalent widths (EWs) of the [OII] doublet in the DESI spectra with the continuum flux predicted from the photometry:
\begin{equation} \label{eqn:ewcorrection}
    F_{\rm [OII]}^{\rm EW-corrected} = (EW_{3726} + EW_{3729}) \cdot F_{3728}^{\rm continuum},
\end{equation}
where $EW_{372\rm{X}}$ is the equivalent width of the line at (rest-frame) 372X $\text{\AA}$ (measured from the \textit{spectrum}) and $F_{3728}^{\rm continuum}$ is the modeled continuum flux around the [OII] doublet at (rest-frame) $\sim$3728 $\text{\AA}$ (inferred from the \textit{photometry}).
All quantities are provided in the \textsc{FastSpecFit} value-added catalog \citep{moustakas23}.

\begin{figure*}
\begin{center}
    \includegraphics[width=0.99\textwidth]{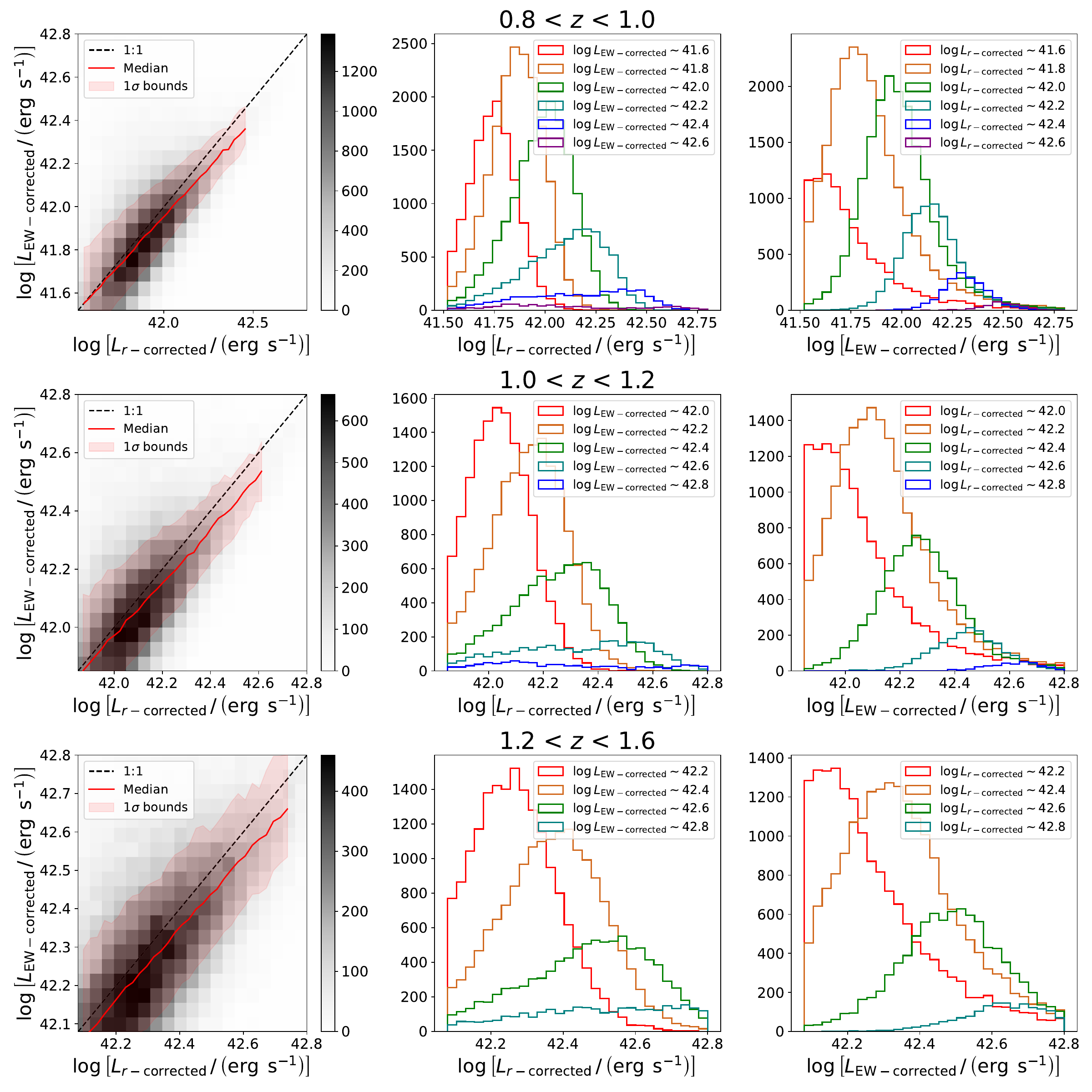}
    \caption{Comparison between aperture corrections. Each row contains one redshift group. \textit{Left column:} equivalent-width-corrected (``EW-corrected'') [OII] luminosity vs. $r$-band-corrected (``$r$-corrected'') [OII] luminosity. Shaded squares and color bar give the number of galaxies in each 2D bin. The one-to-one (black dashed) line traces the median (red solid) within 1 standard deviation and $\sim$0.1 dex, suggesting consistency in the central value. \textit{Middle column:} distribution of $r$-corrected $\log L_{\rm [OII]}$ at fixed EW-corrected $\log L_{\rm [OII]}$. \textit{Right column:} distribution of EW-corrected $\log L_{\rm [OII]}$ at fixed $r$-corrected $\log L_{\rm [OII]}$ (reverse of middle column). }
    \label{fig:aper_comp}
\end{center}
\end{figure*}

Figure~\ref{fig:aper_comp} compares the [OII] luminosities calculated with the $r$-based correction ($L_{r-\rm corrected}$) and the EW-based correction ($L_{\rm EW-corrected}$).
A few conclusions are drawn from Figure~\ref{fig:aper_comp}.
In the left column, the one-to-one line at each redshift traces the median within one standard deviation and $\sim$0.1 dex.
This suggests consistency in the central value between aperture corrections.
However, the middle and right columns show that the distribution of $\log L_{\rm EW-corrected}$ at fixed $\log L_{r-\rm corrected}$ (right column) is more Gaussian than the reverse (middle column).
This result suggests that the $r$-corrected [OII] luminosities are closer to truth.
Further, the EW-based correction relies on a model to predict the continuum from only a few photometric measurements.
This likely introduces more uncertainty than the assumed 1\arcsec{} FWHM seeing of the $r$-based correction.
For these reasons, the $r$-based correction is preferred.

\section{Full Model Constraints} \label{app:corner}
As outlined in Section \ref{sec:model}, we parameterize the \lglg{} distribution of our DESI ELG samples as a function of halo mass.
Doing so allows us to reproduce the projected 2PCFs and number densities of our samples and their dependence on [OII] luminosity and stellar mass.

\begin{figure*}
\begin{center}
    \includegraphics[width=0.99\textwidth]{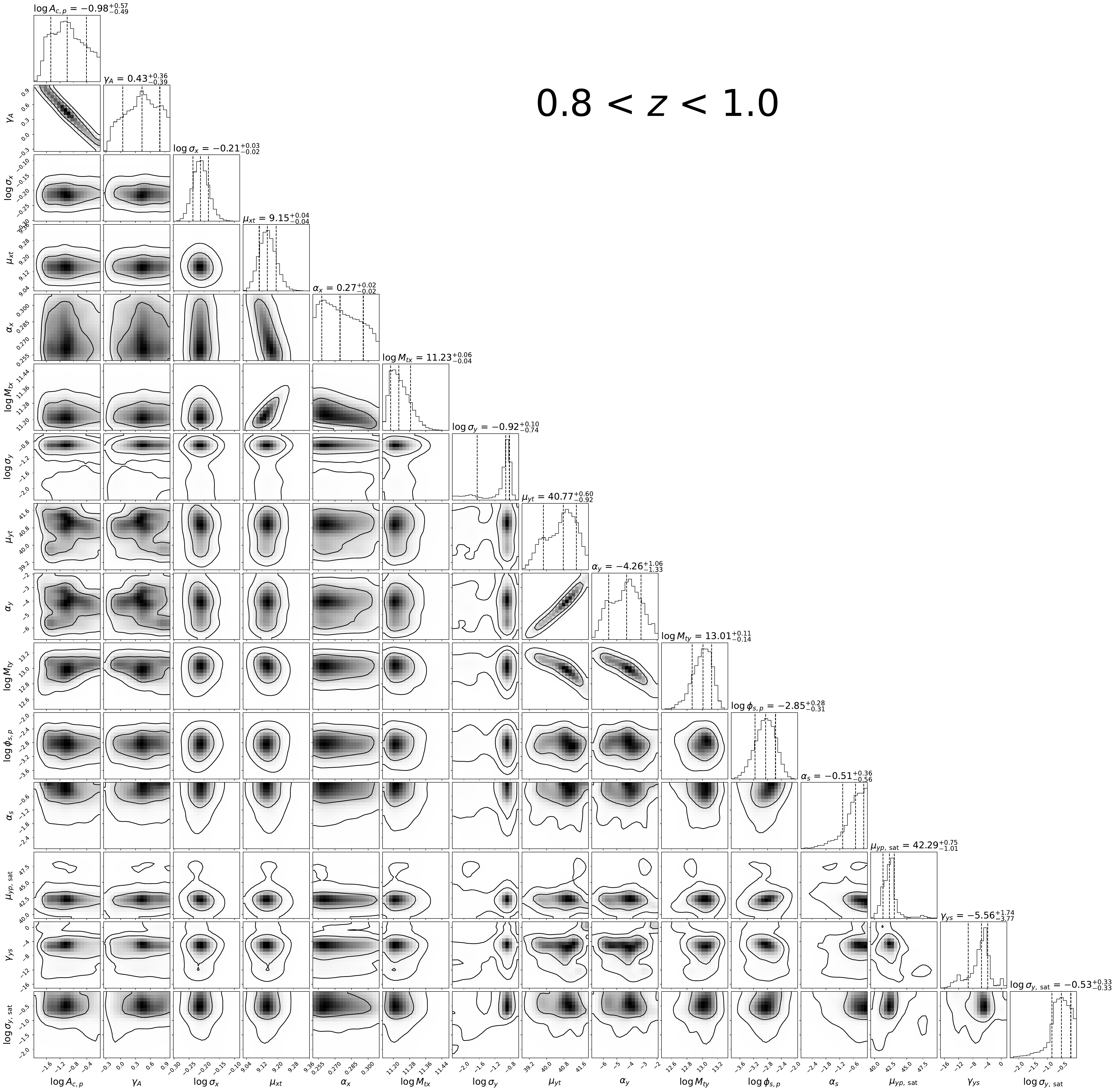}
    \caption{Constraints on all model parameters, for the low-redshift ($0.8<z<1.0$) fitting. In each corner plot, shaded boxes show the binned density of the model posterior. Solid contours enclose the central 68\% and 95\% of the 2D distributions. In each histogram, dashed vertical lines mark the median and central 68\% interval.}
    \label{fig:corner_loz}
\end{center}
\end{figure*}

\begin{figure*}
\begin{center}
    \includegraphics[width=0.99\textwidth]{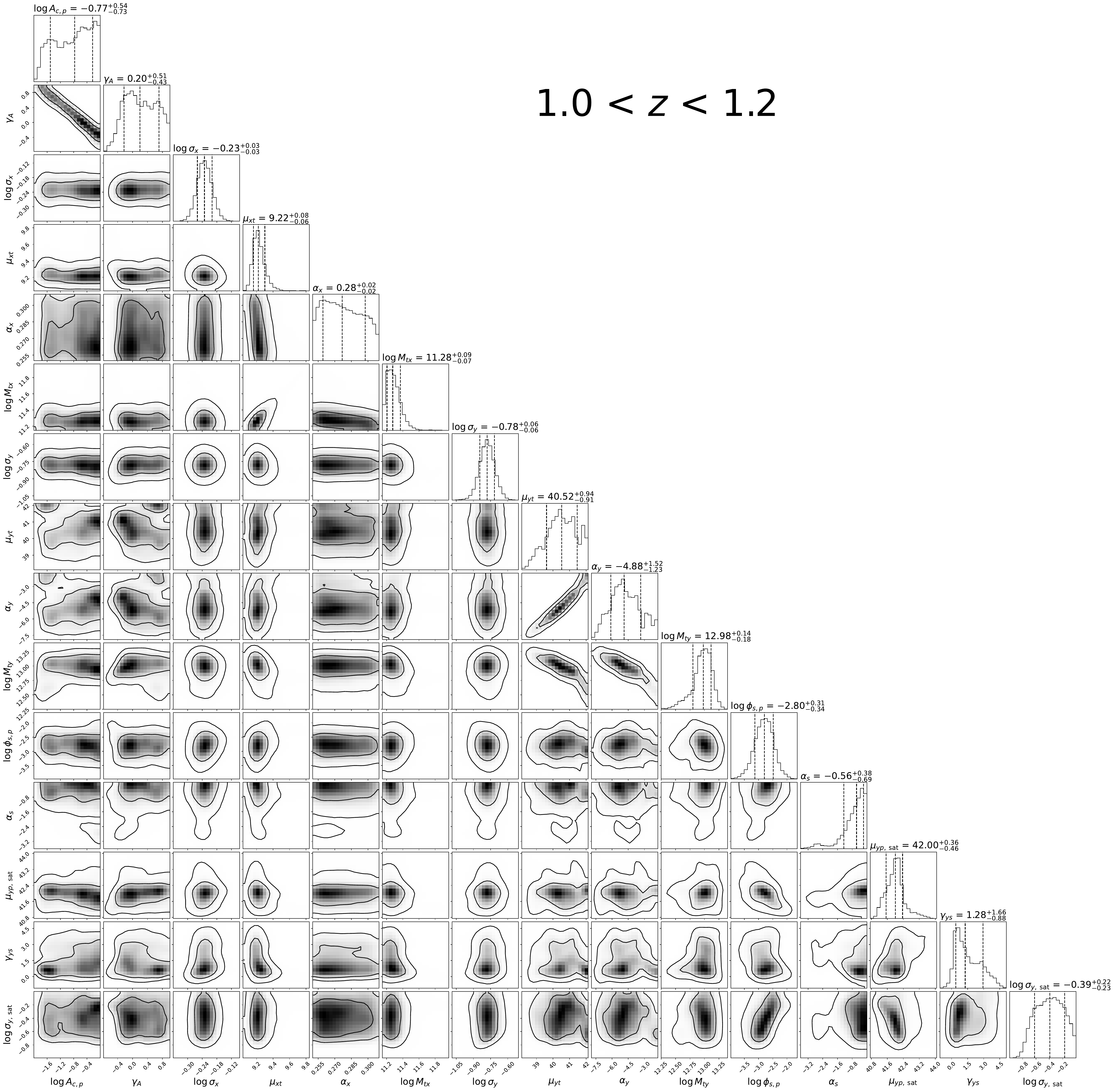}
    \caption{Same as Figure~\ref{fig:corner_loz} but for the $1.0<z<1.2$ intermediate-redshift fitting.}
    \label{fig:corner_midz}
\end{center}
\end{figure*}

\begin{figure*}
\begin{center}
    \includegraphics[width=0.99\textwidth]{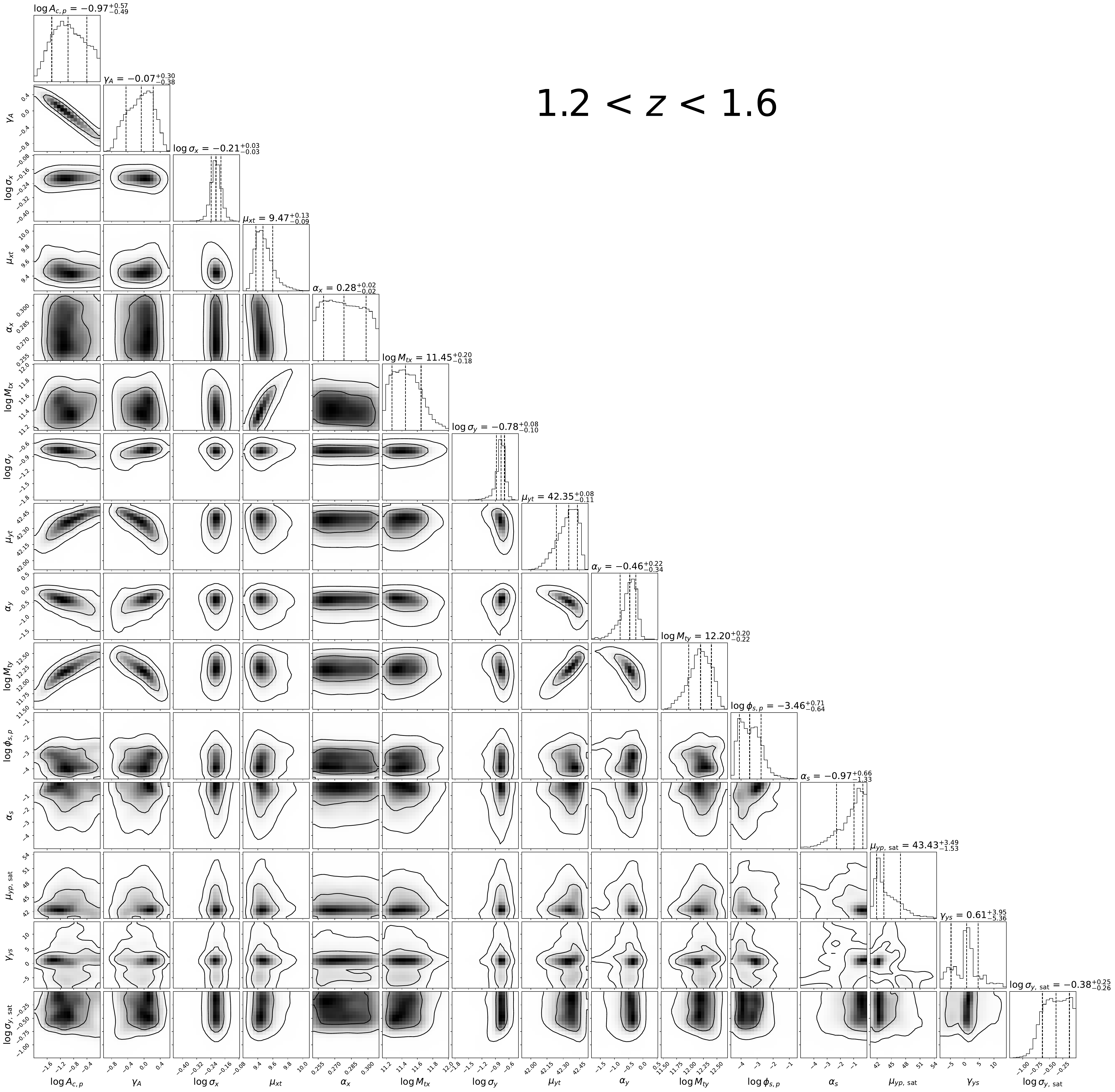}
    \caption{Same as Figure~\ref{fig:corner_loz} but for the $1.2<z<1.6$ high-redshift fitting.}
    \label{fig:corner_hiz}
\end{center}
\end{figure*}

At fixed halo mass, the central galaxy distribution is modeled as a 2D Gaussian.
The satellite distribution is modeled as a modified Schechter function with respect to (logarithmic) stellar mass and a Gaussian distribution of (logarithmic) [OII] luminosity at fixed stellar mass.
Figures~\ref{fig:params_loz}-\ref{fig:params_hiz} show several relations to halo or stellar mass that are reconstructed from a combination of model parameters.
In contrast, Figures~\ref{fig:corner_loz}-\ref{fig:corner_hiz} show the constraints on all model parameters for low, intermediate, and high redshift, respectively.

\end{document}